\documentclass[compsoc, cspaper]{IEEEtran} 
\usepackage{array}
\usepackage{stfloats}
\usepackage{verbatim}
\usepackage{cite}
\usepackage{amsmath,amssymb,amsfonts}
\usepackage{algorithmic}
\usepackage{textcomp}
\usepackage{multirow}
\usepackage[ruled,linesnumbered]{algorithm2e}
\SetKwIF{If}{ElseIf}{Else}{if}{}{else if}{else}{}
\SetKwFor{While}{while}{}{}
\SetKwFor{For}{for}{}{}
\RequirePackage{CJKnumb}
\usepackage{multirow} 
\usepackage{amsmath} 
\usepackage{booktabs,tabularx,makecell}
\usepackage[inline]{enumitem}
\usepackage[table, xcdraw]{xcolor}
\usepackage{subcaption}
\usepackage{scalerel}
\usepackage{tikz}
\usepackage{xcolor}
\usepackage{graphicx}

\usepackage{adjustbox}
\graphicspath{{./images}}

\usetikzlibrary{svg.path}
\definecolor{orcidlogocol}{HTML}{A6CE39}
\tikzset{
  orcidlogo/.pic={
    \fill[orcidlogocol] svg{M256,128c0,70.7-57.3,128-128,128C57.3,256,0,198.7,0,128C0,57.3,57.3,0,128,0C198.7,0,256,57.3,256,128z};
    \fill[white] svg{M86.3,186.2H70.9V79.1h15.4v48.4V186.2z}
                 svg{M108.9,79.1h41.6c39.6,0,57,28.3,57,53.6c0,27.5-21.5,53.6-56.8,53.6h-41.8V79.1z M124.3,172.4h24.5c34.9,0,42.9-26.5,42.9-39.7c0-21.5-13.7-39.7-43.7-39.7h-23.7V172.4z}
                svg{M88.7,56.8c0,5.5-4.5,10.1-10.1,10.1c-5.6,0-10.1-4.6-10.1-10.1c0-5.6,4.5-10.1,10.1-10.1C84.2,46.7,88.7,51.3,88.7,56.8z};
 }
}
\newcommand\orcidicon[1]{\href{https://orcid.org/#1}{\mbox{\scalerel*{
\begin{tikzpicture}[yscale=-1,transform shape]
\pic{orcidlogo};
\end{tikzpicture}
}{|}}}}

\usepackage{hyperref} 
\usepackage{cleveref}

\def\BibTeX{{\rm B\kern-.05em{\sc i\kern-.025em b}\kern-.08em
    T\kern-.1667em\lower.7ex\hbox{E}\kern-.125emX}}
\usepackage{balance}
\begin{document}
\title{Optimizing Multi-DNN Inference on Mobile Devices through Heterogeneous Processor Co-Execution}

\author{Yunquan Gao,
        Zhiguo Zhang,  
        Praveen Kumar Donta,~\IEEEmembership{Senior Member,~IEEE,}\\ 	  
	  Chinmaya Kumar Dehury,
        Xiujun Wang, 
        Dusit Niyato,~\IEEEmembership{Fellow,~IEEE,} 
        Qiyang Zhang
\thanks{Corresponding authors:Zhiguo Zhang, Qiyang Zhang.}
\thanks  
  {
    Y. Gao, X. Wang are with School of Computer Science and Technology, Anhui Engineering Research Center for Intelligent Applications and Security of Industrial Internet, Anhui University of Technology, Ma’anshan, Anhui, 243032, China. E-mail:\{gaoyunquan@bupt.cn; wxj@mail.ustc.edu.cn\};\\  
      Z. Zhang is with School of Computer Science and Technology, Anhui University of Technology, Ma’anshan, Anhui, 243032, China. E-mail:zgzhang@ahut.edu.cn;\\    
        P.K. Donta with Department of Computer and Systems Sciences, Stockholm University, 106 91 Stockholm, Sweden. Email: praveen@dsv.su.se;\\
        C.K. Dehury was with Institute of Computer Science, University of Tartu, Estonia. Now he is with IISER Berhampur, India. Email: dehury@iiserbpr.ac.in;\\
        D. Niyato is with the College of Computing and Data Science, Nanyang Technological University, Singapore 639798. e-mail: dniyato@ntu.edu.sg.\\
        Q. Zhang is with Computer Science School, Peking University, Beijing 100876, China.
  E-mail:qiyangzhang@pku.edu.cn;\\   
		}

}

\markboth{Journal of \LaTeX\ Class Files,~Vol.~14, No.~8, August~2021}%
{Shell \MakeLowercase{\textit{et al.}}: A Sample Article Using IEEEtran.cls for IEEE Journals}

\IEEEtitleabstractindextext{%
 \begin{abstract}
 Deep Neural Networks (DNNs) are increasingly deployed across diverse industries, driving a growing demand to enable their capabilities on mobile devices. However, existing mobile inference frameworks are often rely on a single processor to handle each model's inference, limiting hardware utilization and leading to suboptimal performance and energy efficiency. Expanding DNNs accessibility on mobile platforms requires more adaptive and resource-efficient solutions to meet increasing computational demands without compromising device functionality. Nevertheless, parallel inference of multiple DNNs on heterogeneous processors remains a significant challenge. Several works have explored partitioning DNN operations into subgraphs to enable parallel execution across heterogeneous processors. However, these approaches typically generate excessive subgraphs based solely on hardware compatibility, increasing scheduling complexity and memory management overhead. To address these limitations, we propose an Advanced Multi-DNN Model Scheduling (ADMS) strategy that optimizes multi-DNN inference across heterogeneous processors on mobile devices. ADMS constructs an optimal subgraph partitioning strategy offline, considering both hardware support of operations and scheduling granularity, while employing a processor-state-aware scheduling algorithm that dynamically balances workloads based on real-time operational conditions. This ensures efficient workload distribution and maximizes the utilization of available processors. Experimental results show that, compared to vanilla inference frameworks, ADMS reduced multi-DNN inference latency by 4.04$\times$.
\end{abstract}
}
\maketitle
\IEEEdisplaynontitleabstractindextext

\section{Introduction} \label{sec:introduction}
\IEEEPARstart{I}n recent years, Deep Neural Networks (DNNs) have been widely deployed in various mobile applications such as object detection \cite{liu2020continuous,zhang2023intelligence}, facial recognition \cite{yi2020eagleeye}, and speech recognition \cite{park2018fully}. To reduce interaction latency and lower server-side computing costs, an increasing number of applications are shifting inference tasks to mobile devices.
In many real-world scenarios, multiple independent or related DNN models run concurrently on mobile devices. For instance, in the smart agriculture scenario, farmers capture video frames using smartphone camera and perform real-time parallel inference with multiple DNN models. These models include crop identification~\cite{you2020dnn}, pest and disease detection~\cite{albanese2021automated}, plant health assessment~\cite{demilie2024plant}, and soil quality analysis~\cite{bhat2023soil}.
Consequently, studying real-time multi-DNN collaborative inference on mobile devices is essential for enabling such applications.

However, deploying them on resource-limited mobile devices remains highly challenging due to their complex architectures and the large number of layers and parameters. To address this challenge, extensive research and development efforts have been undertaken by both academia and industry. Nemerous of deep learning (DL) inference frameworks now support mobile platforms\cite{zhang2023comprehensive}, including TensorFlow Lite (TFLite)~\cite{Tensorflow-Lite}, PyTorch Mobile~\cite{pytorch_mobile}, Apple's Core ML~\cite{apple2022coreml}. These frameworks offer mobile-compatible operations required by DNNs to accelerate DNN inference on mobile devices. Nevertheless, the majority of DL inference tasks are performed on CPUs \cite{wu2019machine}. Despite the development of various system of chips (SoCs) on mobile devices~\cite{zhang2023comprehensive}, such as 
Graphics Processing Unit (GPU), Digital signal processing (DSP), and Neural Processing Unit (NPU), which exhibit impressive computational performance, the broad support for inference operators (ops) on mobile CPUs and their gradually improving computational power make them the primary choice for inference tasks. 
Due to the limitations of these accelerators in supporting DNN model operations, the performance of current DL frameworks still has significant room for improvement when performing multi-DNN collaborative execution on mobile devices.

Currently, the literature on DNN inference optimization has limitations in effectively addressing multi-DNN workloads. While some works focus on cloud offloading, others like Wang et al. \cite{wang2021asymo} and Gomatheeshwari et al. \cite{gomatheeshwari2020appropriate} optimize inference for single DNN models only. Although recent approaches like Band \cite{jeong2022band} support multi-DNN inference on heterogeneous processors, they generate excessive subgraphs based solely on hardware support, leading to high memory usage and scheduling complexity. A comprehensive approach that efficiently handles multiple concurrent models while adapting to heterogeneous processor characteristics is still lacking.


To better understand these limitations, we conducted preliminary measurement across 6 different DNN models (ranging from lightweight to complex architectures) on 3 mobile platforms (Redmi K50 Pro, Huawei P20, and Xiaomi 6). Through these experiments, we make the following interesting and useful insights:

1) \textbf{Subgraph partitioning granularity dramatically impacts system resource efficiency and inference performance.} 
Our measurements revealed the excessive subgraphs directly increased inference latency by up to 28\% compared to optimized partitioning, as substantial time was spent on subgraph scheduling and management rather than computation.

2) \textbf{Processor operational status significantly affects sustained inference performance and system stability.} Our thermal stress tests demonstrated that existing frameworks trigger processor throttling after just 2.5 minutes of continuous operation, with CPU frequency dropping from 3GHz to as low as 1GHz and GPU utilization becoming highly unstable. This throttling resulted in up to 4.3$\times$ performance degradation in continuous workloads, emphasizing the importance of processor-state-aware scheduling for maintaining consistent performance in extended inference sessions.

These insights highlight two critical challenges in multi-DNN inference on mobile devices: excessive subgraph fragmentation and performance degradation due to processor state unawareness. To address the first challenge, we introduce an adaptive subgraph partitioning approach that dynamically determines the optimal window size parameter based on both model characteristics and hardware capabilities. This self-tuning mechanism establishes hardware-specific minimum operation requirements for subgraph formation, automatically finding the optimal balance between reducing fragmentation and maximizing processor compatibility for each device-model pair. 
For the second challenge, we develop a comprehensive scheduling strategy that continuously monitors processor operational states, including load, temperature, and frequency. The scheduler integrates this real-time information into a multi-factor priority model that balances deadline urgency, waiting fairness, and resource efficiency in task allocation decisions. By being processor-state-aware, the scheduler ensures balanced workload distribution across heterogeneous processors, prevents throttling, and maintains consistent performance during extended inference sessions.


In this paper, we present a novel Advanced Multi-DNN Model Scheduling (ADMS) framework that constructs and schedules optimal execution plans for collaborative inference on multiple heterogeneous processors by comprehensively considering both DNN model structures and real-time processor status. Extensive experiments demonstrate that ADMS's superior performance in inference latency, energy efficiency, and thermal stability. We summarize our contributions as follows:


\begin{enumerate}
  \item We conduct a comprehensive analysis of multi-DNN inference characteristics on mobile heterogeneous processors, identifying critical performance bottlenecks and quantifying their impact through extensive measurements across various devices and models.
  
  \item We propose an adaptive subgraph partitioning method with hardware granularity control that dynamically determines optimal window sizes for each model-device pair, significantly reducing memory overhead and scheduling complexity while preserving processing efficiency.
  
  \item  We develop a comprehensive multi-factor scheduling algorithm that incorporates real-time processor operational states while balancing deadline urgency, fairness, and resource efficiency. This approach effectively mitigates throttling and ensures consistent performance during extended inference sessions.
  
  \item We design, implement, and evaluate ADMS through extensive evaluations on multiple mobile platforms, demonstrating up to 4.04× reduction in inference latency compared to TFLite and 24.2\% better energy efficiency than Band.
\end{enumerate}

The remainder of the paper is organized as follows: Section \ref{Section:challenges} introduces background concepts and motivation by analyzing challenges in multi-processor inference. Section \ref{section:design} details our system design approach including core components of ADMS. Explains the implementation details with specific algorithms. Section \ref{sec:solution} evaluates the performance through comprehensive experiments on multiple mobile devices. Section \ref{sec:relwork} reviews related works in DNN optimization and processor coordination. Finally, Section \ref{sec:conclusion} concludes the paper and discusses future research directions.


\section{Background And Motivation}\label{Section:challenges}

\subsection{Background}
In DNN model representation, the model is typically structured as a directed acyclic graph (DAG), where each node represents a computational operation, and directed edges indicate dependencies—showing that the output tensor of one operation serves as the input tensor for another. This structure not only defines the execution sequence of operations but also ensures correct data transfer between them. DNN inference primarily consists of two stages: forward propagation and backpropagation, with forward propagation being the core of inference. It processes operations in the order dictated by the DAG, transforming input tensors into output tensors.

\begin{figure}
  \centering
  \includegraphics[width=\linewidth]{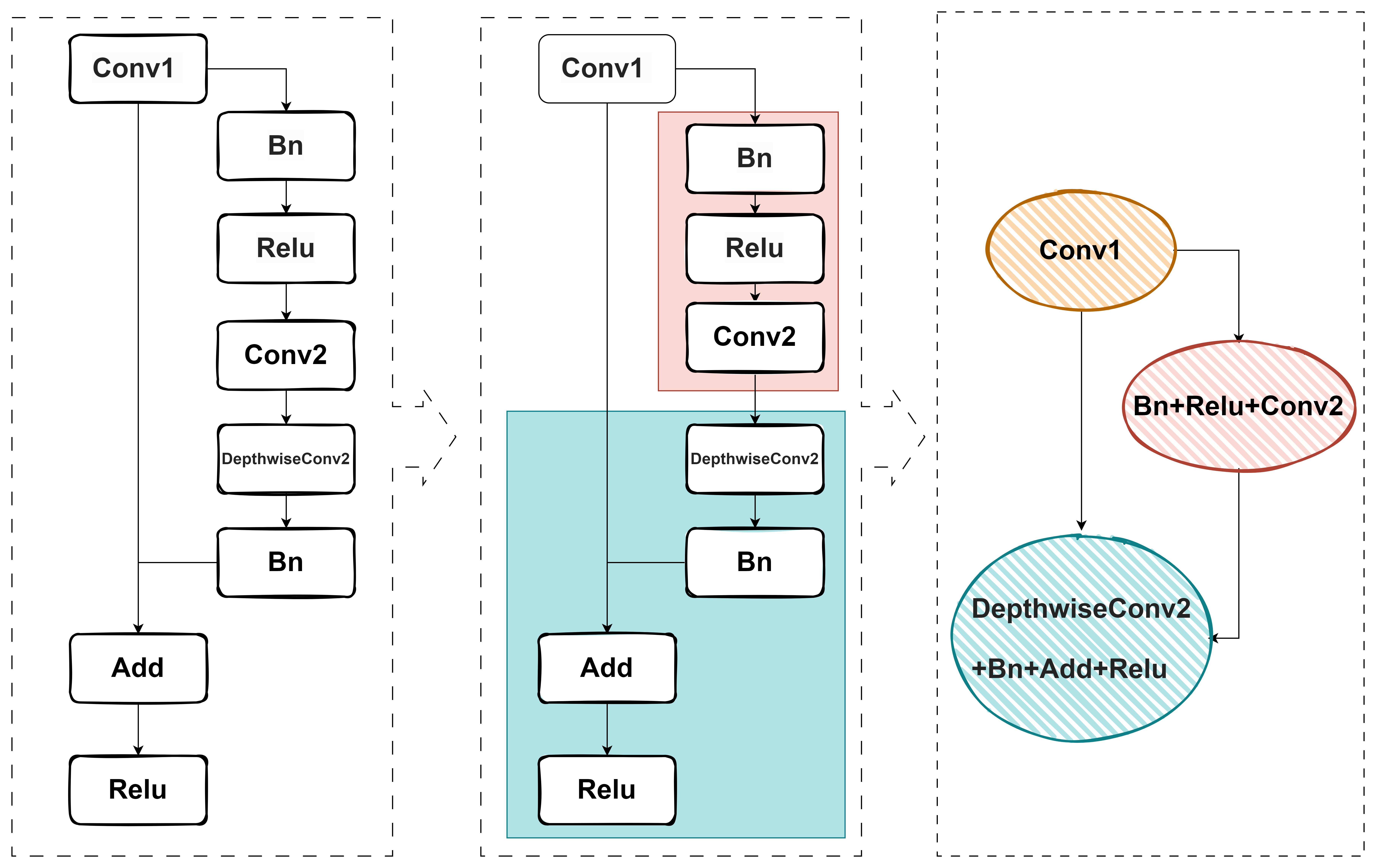}
  \caption{ Subgraph partitioning process for processor-specific execution: original network (left), grouped ops by processor compatibility (middle), and merged processor-specific subgraphs (right).}
  \label{fig:scalesubgraph}
\end{figure}

As shown in \autoref{fig:scalesubgraph}, DNN models on mobile devices are often partitioned into multiple subgraphs to optimize computational efficiency and resource utilization. Each subgraph comprises a set of related ops that can be executed efficiently on a specific processor. The left panel shows the original model structure with individual ops, while the middle panel illustrates how ops are grouped into processor-specific subgraphs (red for GPU-compatible ops, blue for CPU-optimized ops). The right panel demonstrates the fully merged subgraphs that will be scheduled on different processors. This approach helps reduce inference latency and energy consumption by matching ops to their most suitable processor. Current mobile DNN frameworks \cite{zhang2022comprehensive, Tensorflow-Lite, pytorch_mobile, abadi2016tensorflow, apple2022coreml} typically employ static graph partitioning methods based on the DNN models structure, focusing on optimizing the inference latency of a single model.

DNN models encompass a variety of op types, each with varying levels of support across multi-processors on mobile devices. Even within the same processor, the support level for different ops can differ, highlighting the significant heterogeneity of AI accelerators. For instance, \autoref{table:optype} shows the distribution of major op types in eight classic DNN models, with \textit{DeepLabV3} featuring 12 different op types across 134 nodes, illustrating the diversity of ops. This diversity reveals the model's dependency on hardware accelerators, which directly impacts performance.

\begin{figure}
  \centering
  \includegraphics[width=\linewidth]{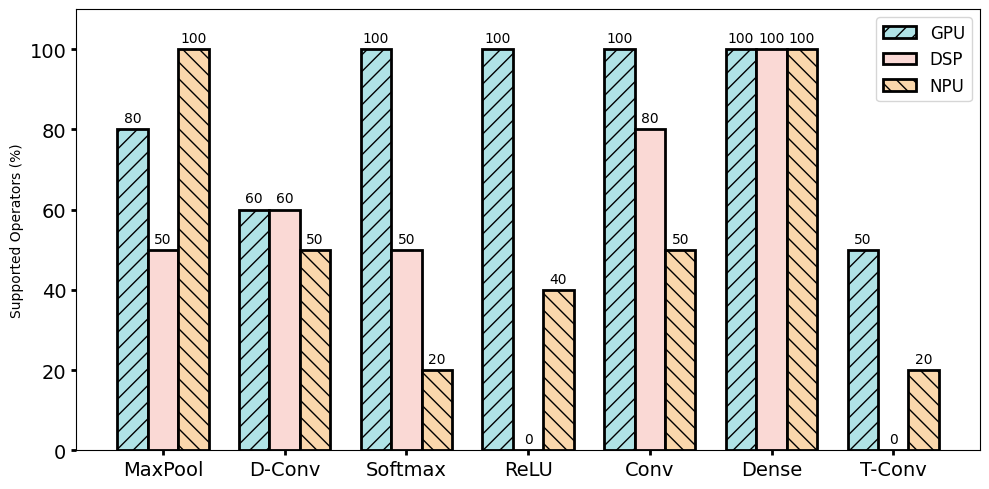}
  \caption{Support for different operation types by various processors on the Redmi K50 Pro.}
  \label{fig:opsupport}
\end{figure}

Hardware heterogeneity is prevalent across various DNN models. As shown in \autoref{fig:opsupport}, the support for different op types varies across processors (such as DSP and NPU) on the Redmi K50 Pro, with many ops not ideally supported. Unfortunately, these hardware limitations cannot be fully mitigate by kernel-level programming on specific processors, mainly because recently developed accelerator core components are often of fixed design, optimized for a limited set of computationally intensive ops. For instance, Google's Edge TPU\cite{google2020edgeTPU} uses a systolic array, while Huawei's Da Vinci architecture uses 3D cube technology, both optimized for specific ops with defined parameters.

\begin{table}[t]
  \centering
  \caption{Proportional distribution of operation types in DNN models.}
  \begin{tabular}{@{}lccccc@{}} 
    \toprule
    \textbf{Model Names} & \multicolumn{5}{c}{\textbf{Operation Types}} \\
    \cmidrule(lr){2-6}
     & \textbf{ADD(\%)} & \textbf{C2D(\%)} & \textbf{DLG(\%)} & \textbf{DW(\%)} & \textbf{Others(\%)}\\ 
    \midrule
    \textbf{Arcface} & 15.28 & 48.61 & 1.39 & 23.61 & 6.94 \\
    \textbf{DeepLabV3} & 14.93 & 28.36 & 16.42 & 12.69 & 7.46 \\
    \textbf{East} & 14.16 & 55.75 & 4.42 & - & 10.62 \\
    \textbf{EfficientNet4} & 18.85 & 50.0 & 1.64 & 24.59 & 1.64 \\
    \textbf{HandLmk} & 23.75 & 48.28 & - & 23.75 & 3.45 \\
    \textbf{ICN} & 26.83 & 57.32 & 6.1 & 2.44 & 4.88 \\
    \textbf{InceptionV4} & - & 69.3 & 9.3 & - & 20.47 \\
    \textbf{MobileNetV2} & 14.71 & 52.94 & 2.94 & 25.0 & 1.47 \\
    \bottomrule
  \end{tabular}
  \label{table:optype}
\end{table}

\subsection{Preliminary Measurement and Insights}
After reviewing related work, we now examine the specific characteristics of DNN inference on mobile devices. To derive execution strategies and scheduling characteristics for DNN model inference, we conducted measurements using TFLite on Huawei P20 (Kirin 970) and Redmi K50 Pro (Dimensity 9000).

We deployed two popular DNN models on these devices, namely MobileNet~\cite{howard2017mobilenets} and EfficientDet~\cite{tan2020efficientdet}. MobileNet significantly reduces model parameters and computation by using depthwise separable convolutions, decomposing standard convolutions into depthwise and pointwise convolutions, making it highly suitable for real-time mobile applications. In contrast, EfficientDet has a much more complex op structure, combining EfficientNet and BiFPN architectures for efficient multi-scale feature fusion, providing excellent object detection performance while maintaining computational efficiency. 

\subsubsection{Inference Characteristics}
\begin{figure}[!t]
  \centering
  \subcaptionbox{\textbf{MobileNetV1 model}\label{fig:subfig1}}{
  \includegraphics[width=0.8\columnwidth]{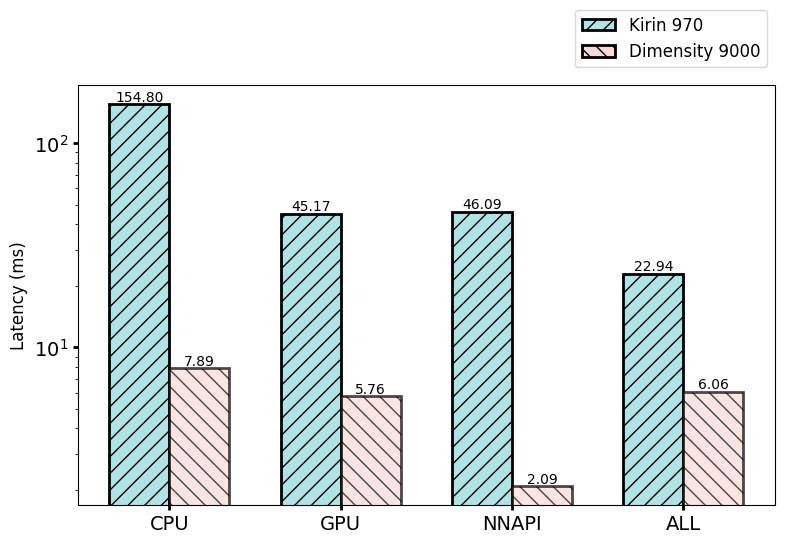}}
  
  
  \subcaptionbox{\textbf{EfficientDet model}\label{fig:subfig2}}{
  \includegraphics[width=0.8\columnwidth]{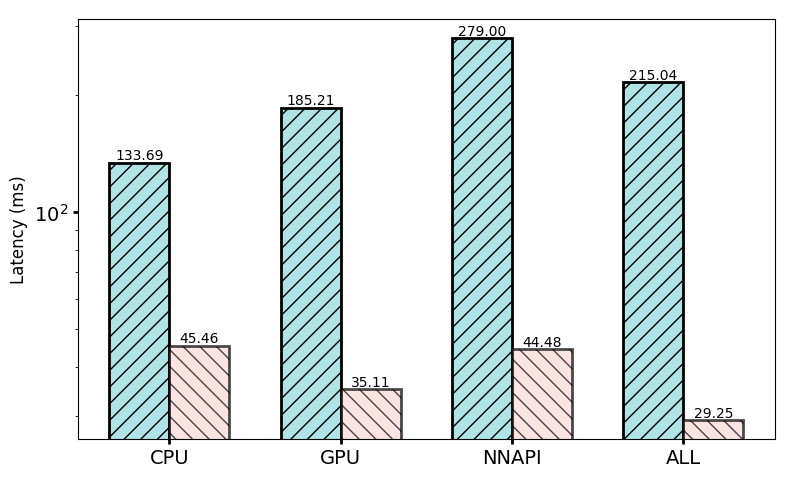}}
  \caption{Average latency of DNN inference on single and multi-processors on the Android platform using Kirin 970 and Dimensity 9000 Chipsets.}
  \label{fig:doublelatency}
\end{figure}

Through the analysis of the measurement results, we identified several key insights:

\textbf{Besides the CPU, other processors also exhibit impressive performance in executing DNN inference.} As shown in \autoref{fig:doublelatency}, when executing the lightweight MobileNet model on Dimensity 9000, the execution efficiency driven by NNAPI~\cite{google_nnapi} on DSP and NPU is 3$\times$ faster than on the CPU alone, with the GPU also showing stronger inference capabilities than the CPU. This performance improvement continually increases with the evolution of SoCs, with the speedup on Dimensity 9000 reaching up to 23$\times$ when executing MobileNet tasks. Therefore, fully utilizing  different processor capabilities for  collaborative inference is crucial.

  \textbf{Multi-processor inference is not always the ideal situation.} As the experimental results in \autoref{fig:doublelatency} show, while the efficiency of performing DNN inference using a single processor is lower than that of collaborative inference across multi-processors, in some specific cases, the latency of multi-processor inference is actually higher than that of a single processor. For example, executing EfficientDet on Kirin 970 and MobileNet on Dimensity 9000 shows higher latency with multi-processor inference. This is mainly because TFLite lacks an effective collaborative scheduling mechanism to fully utilize processor  capabilities during multi-processor inference.

  \textbf{Fallback ops significantly impact the efficiency of DNN model inference.} By tracking the scheduling process of models executed on multi-processors, we found that when ops in the model are assigned to processors that do not support their execution, these ops default to being rerouted to the CPU, even if other processors could execute these ops more efficiently. These unsupported ops vary among different SoCs and are more pronounced on older SoCs. On Kirin 970, for example, the presence of numerous fallback ops leads to massive tensor transfer costs, resulting in lower multi-processor performance than single CPU performance. 
  

\subsubsection{The performance of Multi-Processor Inference}
Through further analysis of our measurement data, we identified several critical challenges that impact the performance of multi-DNN inference on heterogeneous processors:

\textbf{Concurrent model execution reveals scheduling inefficiencies in heterogeneous environments.} As shown in \autoref{table:processorlatency}, when multiple models are processed simultaneously, the performance degradation is non-uniform across different processors. While some processors like MediaTek NPU show modest latency increases (27\% when increasing from 1 to 4 concurrent models), others like the Hexagon 682 DSP experience dramatic performance collapse (1202\% increase when running 4 concurrent models compared to a single model). This disparity highlights significant differences in how current frameworks manage resource sharing across heterogeneous processors, revealing opportunities for substantial performance gains through more intelligent workload distribution.

\renewcommand{\arraystretch}{1.2}

\begin{table}[t]
  \centering
  \caption{Impact of parallel inference on latency: average latency (ms) for MobileNetV1 on various processors with different levels of concurrency.}
  \resizebox{.95\columnwidth}{!}{
  \begin{tabular}{>{\centering\arraybackslash}m{2.5cm} >{\centering\arraybackslash}m{2.5cm} >{\centering\arraybackslash}p{1.5cm} >{\centering\arraybackslash}p{1.5cm} >{\centering\arraybackslash}p{1.5cm}}
  \toprule
  \textbf{Device Names} & \textbf{Accelerators} & \multicolumn{3}{c}{\textbf{The number of concurrently inferred models}} \\ \cmidrule(lr){3-5}
  & & \textbf{1} & \textbf{2} & \textbf{4} \\ \midrule
  \multirow{3}{*}{\textbf{Redmi K50 Pro}} & \textbf{Mali-G710 GPU} & 3.65 & 7.88 & 9.09 \\ \cmidrule(lr){2-5}
  & \textbf{MediaTek APU 5.0} & 8.24 & 10.71 & 16.97 \\ \cmidrule(lr){2-5}
  & \textbf{MediaTek NPU} & 1.88 & 2.13 & 2.39 \\ \midrule
  \multirow{2}{*}{\textbf{HuaWei P20}} & \textbf{Mali-G72 GPU} & 45.35 & 76.77 & 114.88 \\ \cmidrule(lr){2-5}
  & \textbf{Dual-core NPU} & 70.15 & 220.07 & 429.1 \\ \midrule
  \multirow{2}{*}{\textbf{Xiaomi 6}} & \textbf{Adreno 540 GPU} & 7.89 & 7.96 & 8.1 \\ \cmidrule(lr){2-5}
  & \textbf{Hexagon 682 DSP} & 46.77 & 277.14 & 609.44 \\ \bottomrule
  \end{tabular}}%
  \label{table:processorlatency}%
  \end{table}


\textbf{Subgraph partitioning granularity dramatically impacts system resource efficiency and inference performance.} Our experiments show that different models produce vastly different numbers of subgraphs under current partitioning strategies. As documented in \autoref{table:subgraph_counts}, lightweight models like East generate only 4 total subgraphs on the Redmi K50 Pro, while complex models like DeepLabV3 produce over 3,300 subgraphs. This excessive fragmentation consumes disproportionate memory resources and complicates scheduling decisions, directly impacting inference efficiency. When measuring the relationship between subgraph count and inference performance, we observed that reducing subgraph count through optimized partitioning can decrease inference latency by up to 28\%.

\begin{figure*}[t]
\centering
\includegraphics[width=\textwidth, height=0.5\textheight, keepaspectratio]{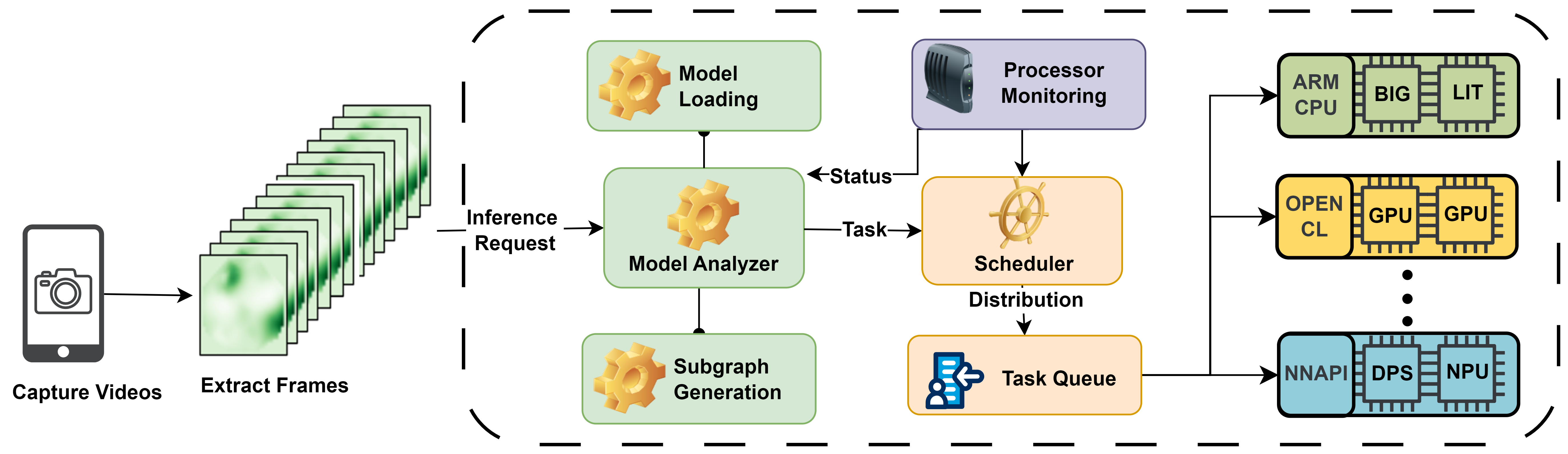}
\caption{The overview of ADMS.}
\small{The workflow shows video frame extraction (left), the three core components - Model Analyzer (green), Scheduler (orange), and Processor Monitor (purple) - and heterogeneous processor allocation (right).}
\label{fig:overview}
\end{figure*}




\textbf{Processor operational status significantly affects sustained inference performance and system stability.} Our measurements revealed that inference performance is not static but degrades over time under sustained workloads. When monitoring processor state during continuous operation, we observed that CPU utilization patterns shift significantly after approximately 150 seconds of operation, with frequency fluctuations increasing by 217\% and temperature rising by 18°C on average. These state changes directly correlate with inference speed reductions, as the system attempts to manage thermal constraints by throttling performance. This dynamic behavior is not accounted for in current scheduling approaches, which maintain static allocation patterns regardless of changing processor conditions.

These observations highlight the need for a fundamentally different approach to multi-DNN inference on mobile devices—one that adaptively manages subgraph granularity, intelligently distributes concurrent workloads through processor-state-aware scheduling, and responds to changing processor conditions during extended operation.

\section{System Design}
\label{section:design}

To address these challenges, we propose the ADMS, a novel approach that optimizes multi-DNN inference across heterogeneous processors through fine-grained subgraph partitioning and dynamic scheduling.


\subsection{System Overview}
ADMS is specifically designed for mobile processors and DNN inference, adopting a multi-DNN collaborative inference architecture supported by heterogeneous processors. The system consists of three core components working in tandem to optimize performance:
1) A Model Analyzer that addresses the subgraph partitioning challenge by dividing DNN models into optimally-sized subgraphs with appropriate hardware support \cite{jeong2022band, jia2022codl}, reducing memory usage and scheduling complexity;
2) A Scheduler that tackles processor contention through fine-grained scheduling plans that coordinate DNN execution across multiple processors and efficiently manage fallback operations \cite{kang2021lalarand, xiang2019pipelined}; and
3) A Hardware Monitor that mitigates thermal throttling issues by continuously tracking processor status \cite{tan2024thermal}, providing real-time data to support optimized scheduling decisions.

During system operation, as illustrated in the object detection application case in \autoref{fig:overview}, the application converts video content into image frames, which are then processed by the model analyzer via the model loader. The analyzer decomposes the DNN model into multiple subgraphs to facilitate independent inference based on the hardware support of the operations. This step aims to alleviate computational load on a single processor and improve processing efficiency. The dynamic scheduler determines the execution order and resource allocation of the subgraphs based on real-time data from the hardware monitor, including processor load, temperature, and operational status. Real-time monitoring by the hardware monitor ensures that the scheduler can formulate optimal scheduling plans, optimizing the processing of subgraphs in the task queues for efficient execution. Once all subgraph tasks are completed, the entire model inference task is finished, and the system returns the results to the application.Through the collaborative efforts of these components, the ADMS can efficiently handle dynamic mobile workload changes while maximizing resource utilization \cite{fan2023sparse, wang2018optic}.

\subsection{Model Analyzer}
The Model Analyzer addresses the challenges of subgraph partitioning and fallback ops by intelligently decomposing DNN models into hardware-appropriate subgraphs. It comprises two main modules: the model loader for importing DNN models, and the subgraph generator for parsing these models into independently inferable subgraphs based on hardware support.

\begin{figure}[!htbp]
    \centering
    \subcaptionbox{\textbf{An example model demonstrates varying op support.}\label{fig:analyze1}}{
    \includegraphics[width=.95\columnwidth]{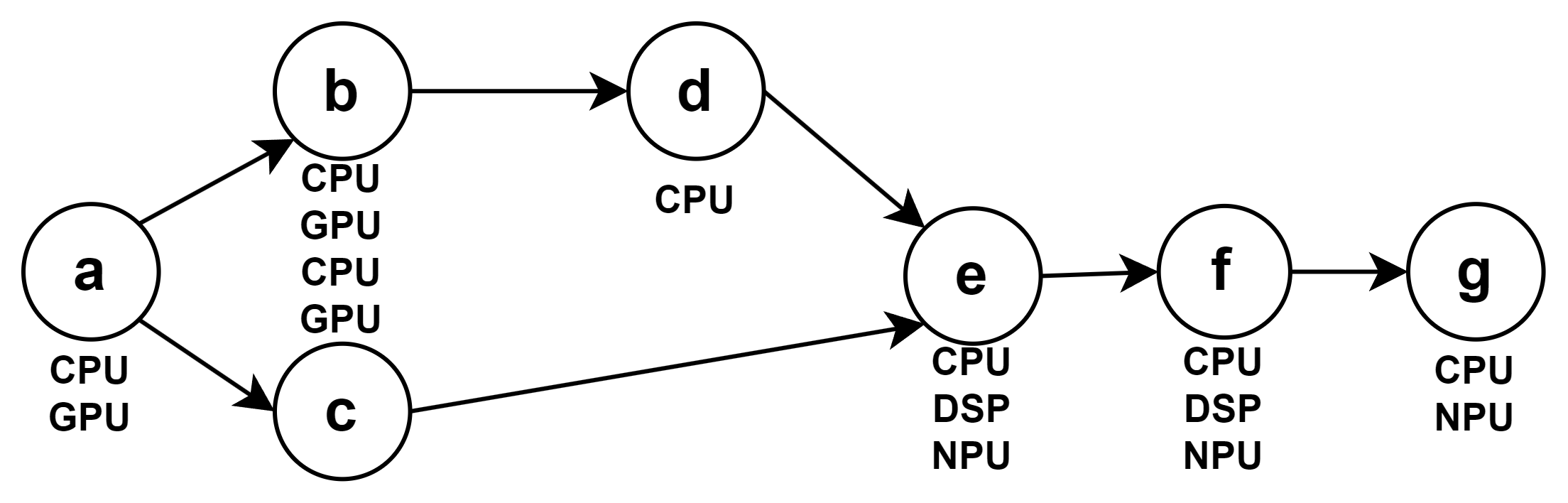}}
    \subcaptionbox{\textbf{For GPUs, TensorFlow Lite creates only a single execution schedule.}\label{fig:analyze2}}{
    \includegraphics[width=.95\columnwidth]{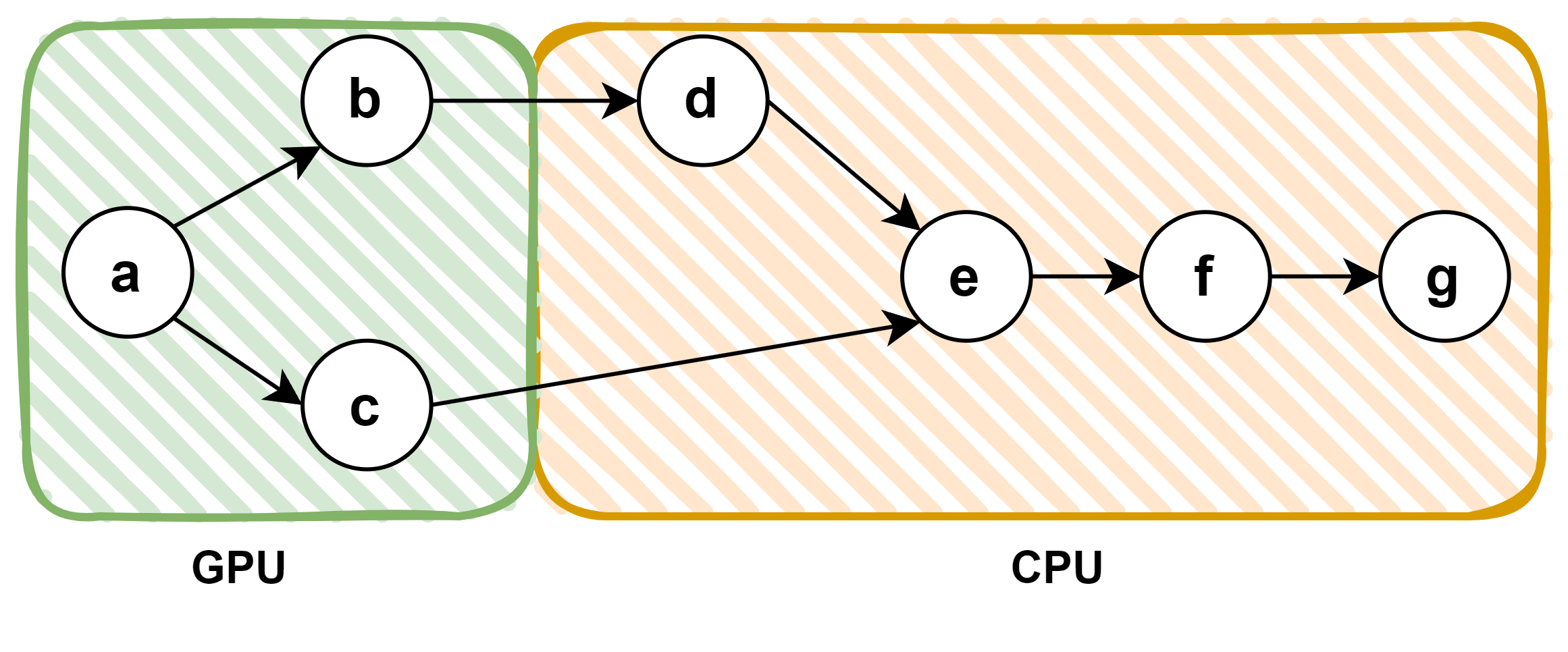}}
    \subcaptionbox{\textbf{Grouping ops into unit subgraphs.}\label{fig:analyze3}}{
    \includegraphics[width=.95\columnwidth]{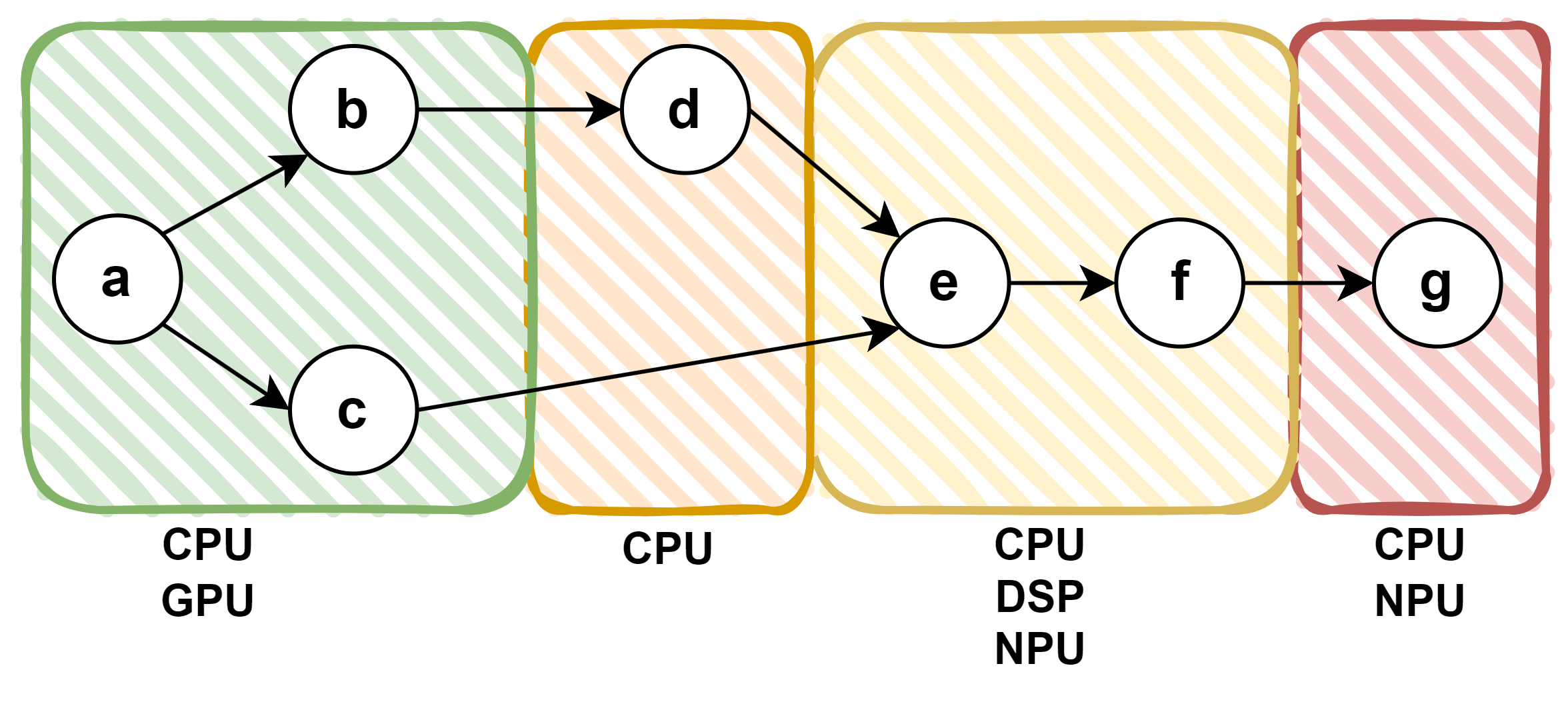}}
    \caption{Sample diagram for subgraph generation.}
    \label{fig:subgraphger}
\end{figure}

As shown in Figure \ref{fig:subgraphger}(a), each op in a DNN model is represented as a node with directed edges indicating dependencies. Traditional op-level scheduling becomes exponentially complex as models grow (e.g., seven nodes can have 72 possible scheduling combinations). Current frameworks typically execute all ops on a single processor, with unsupported ops falling back to the CPU. Figure \ref{fig:subgraphger}(b) illustrates this limitation where only three of seven ops run on the GPU while the rest fall back to the CPU, resulting in substantial tensor transfers and increased latency.ADMS employs a two-step approach to generate efficient subgraphs:

\textbf{Unit Formation}: The analyzer first groups adjacent ops with consistent processor support. As shown in Figure \ref{fig:subgraphger}(c), ops with common hardware support form a unit, while ops with different support requirements form separate units.

\textbf{Subgraph Construction}: Adjacent units are then merged, but only when they share common processor support. This optimizes resource utilization while maintaining execution efficiency.

\begin{table}[t]
    \centering
    \caption{Subgraph and Operation Counts for Various Models on Redmi K50 Pro.}
    \resizebox{.95\columnwidth}{!}{
    \begin{tabular}{
        >{\centering\arraybackslash}m{3cm}
        >{\centering\arraybackslash}m{2cm}
        >{\centering\arraybackslash}m{1.5cm}
        >{\centering\arraybackslash}m{1.5cm}
        >{\centering\arraybackslash}m{1.5cm}
    }
      \toprule
      \textbf{Model names} & \textbf{Operations} & \multicolumn{3}{c}{\textbf{Subgraph Counts}} \\
      \cmidrule(lr){3-5}
      & & \textbf{Unit} & \textbf{Merged} & \textbf{Total} \\ 
      \midrule
      \textbf{East} & 108 & 1 & 0 & 4 \\ 
      \textbf{YoloV3} & 232 & 2 & 3 & 9 \\ 
      \textbf{MobileNetV1} & 31 & 4 & 24 & 42 \\ 
      \textbf{MobileNetV2} & 66 & 26 & 860 & 968 \\ 
      \textbf{ICN\_quant} & 77 & 33 & 1496 & 1644 \\ 
      \textbf{DeepLabV3} & 112 & 65 & 3076 & 3329 \\ 
      \bottomrule
    \end{tabular}}%
    \label{table:subgraph_counts}%
  \end{table}

\begin{figure}[t]
  \centering
  \includegraphics[width=\linewidth]{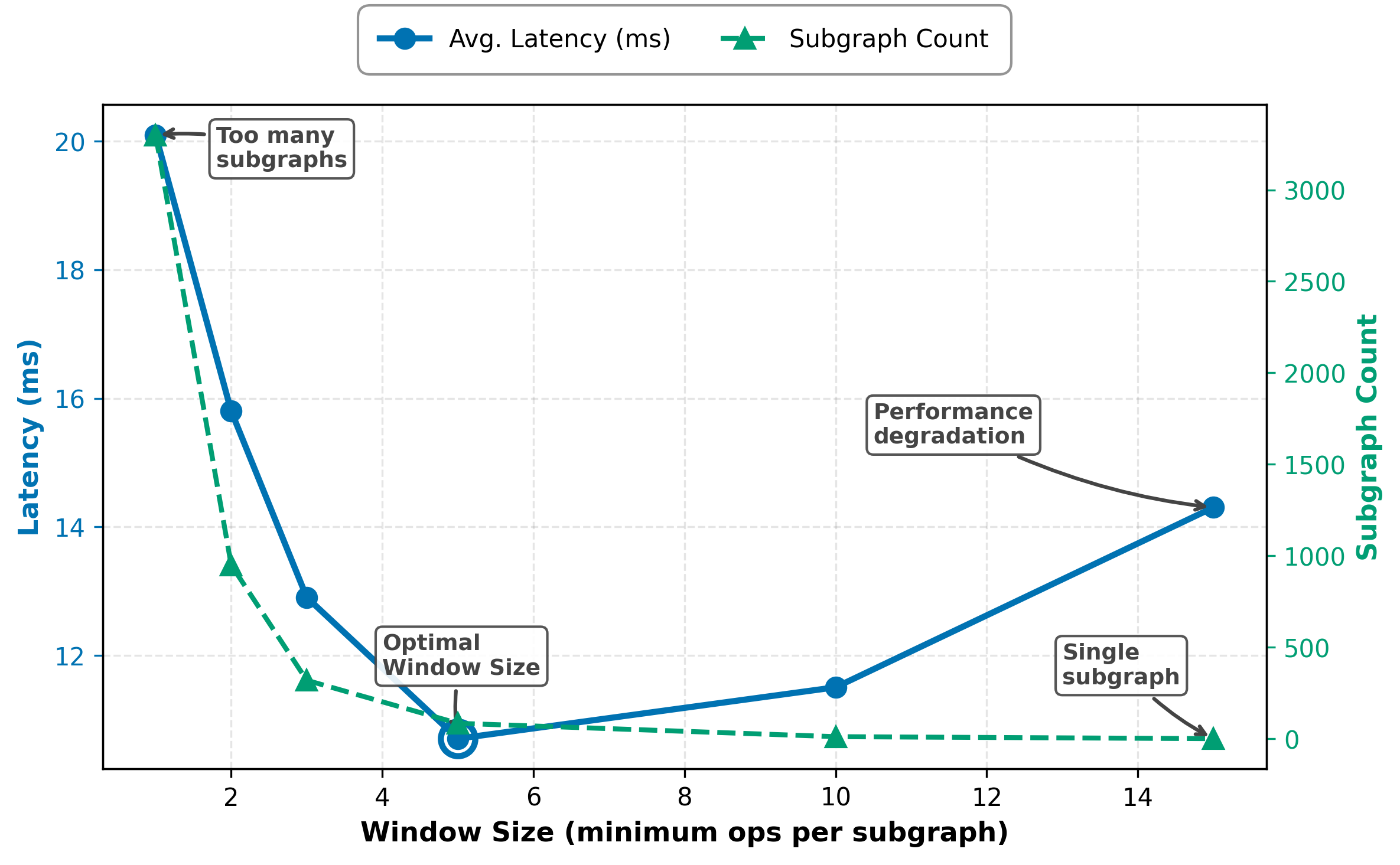}
  \caption{Relationship between window size, inference latency, and subgraph count for DeepLabV3 on Redmi K50 Pro, highlighting the optimal balance at window size 5.}
  \label{fig:windows}
\end{figure}

While Band \cite{jeong2022band} uses similar methods, we found that generating subgraphs solely based on hardware support can produce an excessive number of subgraphs on resource-constrained devices. As shown in \autoref{table:subgraph_counts}, models with low hardware support like DeepLabV3 can generate up to 3329 subgraphs, significantly straining the system. To address this, we introduced the \texttt{window\ size(ws)} parameter defining the minimum ops required to create a subgraph on a target processor. Our experiments (Figure \ref{fig:windows}) show that increasing \texttt{ws} significantly decreases subgraph latency and improves frame rate, while rapidly reducing the number of subgraphs—eventually to a single consolidated graph at the highest settings. However, excessively large values can negatively impact performance. For each model-processor combination, we empirically determine the optimal \texttt{ws} configuration and store it for runtime use, balancing subgraph count reduction and inference performance.



The core implementation of our subgraph generation is detailed in Algorithm~\ref{alg:get_unit_subgraphs}, which efficiently identifies subgraphs with minimal fallback ops. The algorithm begins by determining the number of available processors and clearing existing unit subgraphs (Lines 1-2). For processors with no fallback ops, it creates unit subgraphs containing the entire set of ops (Lines 3-7). For processors requiring fallback handling, it identifies unsupported ops for each processor, builds a support table (Lines 9-17), and resolves optimal subgraph configurations (Lines 18-20). This algorithm minimizes fallback operations by analyzing processor compatibility and constructing optimized subgraphs, enhancing both adaptability and performance across various mobile devices. Our window size optimization strategy, combined with this efficient algorithm implementation, makes ADMS a robust solution for multi-DNN inference on resource-constrained devices.

\begin{algorithm}[t]
    \caption{Get Unit Subgraphs}
    \label{alg:get_unit_subgraphs}
    \KwIn{monitor, model\_profile, subgraph\_config}
    \KwOut{unit\_subgraphs, status}
    num\_processor $\gets$ monitor.GetNumProcessors()\;
    Clear unit\_subgraphs\;
    \If{NOT NeedFallbackSubgraph}{
    entire\_ops $\gets$ \{0, 1, ..., model\_profile.num\_ops - 1\}\;
    \For{each processor\_id from 0 to num\_processor - 1}{
    \If{IsProcessoValid(processor\_id)}{
    unit\_subgraphs.push\_back(processor\_id, entire\_ops, \{0\})\;
    }
    }
    }
    \Else{
    op\_sets\_ignore $\gets$ \{\}\;
    \For{each processor\_id from 0 to num\_processor - 1}{
    processor\_op\_sets $\gets$ GetSubgraphsForFallbackOps(processor\_id)\;
    \For{each processor\_and\_ops in processor\_op\_sets}{
    \If{processor\_and\_ops.op\_indices.size $<$ subgraph\_config.windows\_size}{
    op\_sets\_ignore[processor\_id].insert(\\
    processor\_and\_ops.op\_indices)\;
    }
    }
    }
    op\_support\_table $\gets$ BuildOpSupportTable(model\_profile,\\
    num\_processor, op\_sets\_ignore)\;
    unit\_subgraphs $\gets$ ResolveSubgraphs(op\_support\_table,\\
    num\_processor)\;
    }
    status $\gets$ ValidateAndSetUnitSubgraphs(unit\_subgraphs, model\_profile)\;
    \Return unit\_subgraphs, status\;
\end{algorithm}

\subsection{Hardware Monitor}

The Hardware Monitor component addresses processor thermal throttling by providing real-time insights into the device's operational state, enabling proactive task allocation to prevent performance degradation. It monitors critical metrics including temperature, processor availability, utilization rates, power consumption, and workloads across all inference accelerators (CPU, GPU, DSP, and NPU). Temperature monitoring is particularly crucial, as excessive heat can trigger processor throttling or shutdown, while battery status tracking helps balance performance needs against power constraints, especially during extended inference sessions. Since the Android SDK cannot directly access these system metrics, we developed a dedicated monitoring program that efficiently retrieves this information from system files. This program employs optimized sampling frequencies that balance monitoring accuracy with minimal overhead, providing the scheduler with timely data to make informed allocation decisions while maximizing resource utilization between heterogeneous processors.

This program obtains the device's temperature information by reading files in the '\texttt{/sys/devices/virtual/thermal/}' directory and accessing the '\texttt{/sys/devices/system/cpu/}' directory to gather processor availability status, including each CPU core's online status, frequency, and usage. It obtains GPU status information via OpenGL interfaces, while it retrieves DSP and NPU information from NNAPI interfaces. Additionally, we used the Android SDK to gather battery status and memory usage to aid in scheduling decisions.

The program employs multithreading and a caching mechanism to reduce the frequency of file reads, with the entire data retrieval process taking approximately 10 ms, ensuring real-time accuracy of scheduling decisions. This caching strategy significantly improves performance by storing frequently accessed hardware status information in memory, eliminating redundant file system operations that would otherwise introduce latency of 40-50 ms per request. The cache is refreshed at carefully tuned intervals to maintain data accuracy while minimizing I/O overhead. By comprehensively understanding the mobile device's status through this method, we provide accurate data support to the model parser and dynamic scheduler, optimizing scheduling strategies and enhancing system performance. This mechanism ensures that our scheduling system operates in the best possible state, effectively improving the performance and responsiveness of DNN applications.

\subsection{Scheduler}



The Scheduler resolves processor contention by dynamically coordinating task execution across heterogeneous processors based on real-time status information. When a mobile application requests DNN model inference for the first time, the scheduler sends the model to the analyzer for evaluation. The generated subgraphs are stored in a configuration file for future use. For previously analyzed models, inference requests go directly to the scheduler, which retrieves the subgraph information and adds tasks to the execution queue.

\begin{figure}[t]
  \centering
  \includegraphics[width=\linewidth]{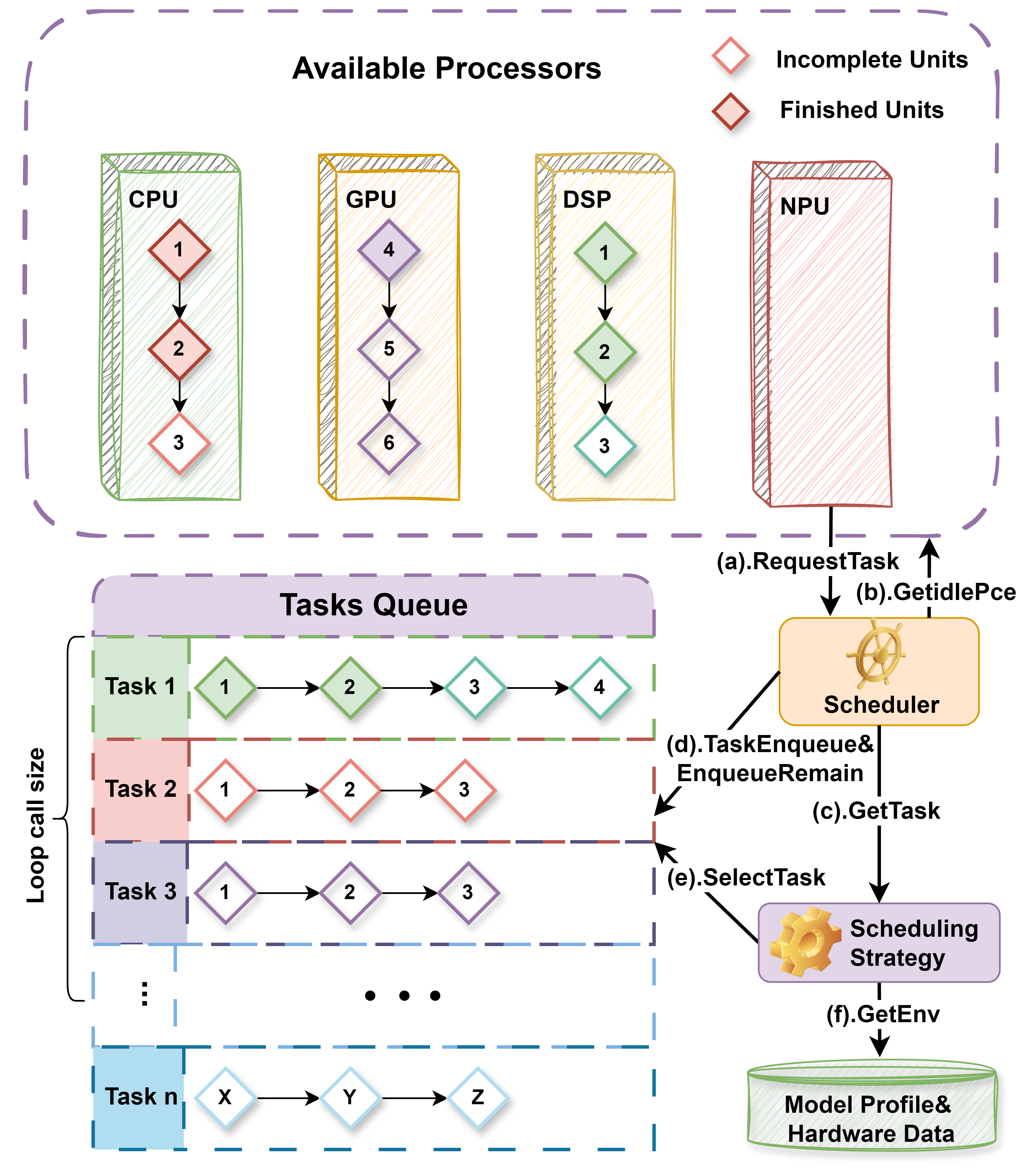}
  \caption{ADMS scheduler workflow illustrating the key operations: processor task requests, idle processor detection, task queue management, and priority-based scheduling using model profiles and hardware status data.}
  \label{fig:Scheduler}
\end{figure}

Our scheduling approach for heterogeneous multi-processor environments addresses fundamental challenges identified in classical scheduling theory \cite{liu1973scheduling, shirazi1995scheduling}. Traditional schedulers often fail in mobile DNN contexts due to varying hardware capabilities, thermal constraints, and dynamic workloads. Building upon established scheduling principles while adapting them for mobile DNN inference, we derive a priority-based model that balances three critical scheduling factors.

Deadline urgency represents the first factor in our scheduling model. SLO define acceptable response times for tasks, and our scheduler prioritizes tasks approaching their deadlines following principles established in real-time systems \cite{stankovic1998deadline}. We quantify this urgency as:
\begin{equation}
S_{deadline} = \gamma (T_{SLO} - T_{latency})
\end{equation}
where $T_{SLO}$ represents the expected response time and $T_{latency}$ is the estimated execution latency. This approach extends traditional deadline-based scheduling by using service-level agreements rather than hard deadlines.

Waiting fairness constitutes the second essential factor in our model. Drawing from fair queuing algorithms \cite{demers1989analysis}, we normalize waiting time by task complexity to prevent starvation of complex tasks:
\begin{equation}
S_{wait} = -\alpha \frac{T_{current} - T_{enqueue}}{T_{avg}}
\end{equation}
where $T_{enqueue}$ is when the task entered the queue, $T_{current}$ is the current time, and $T_{avg}$ is the average execution time.

Resource efficiency forms the third component of our scheduling model. Extending processor affinity concepts \cite{anderson2000early}, we account for processor load and task complexity:
\begin{equation}
S_{resource} = \delta \left(\frac{2B_{current} - B_{max}}{B_{max}}\right) C_{remaining}
\end{equation}
where $B_{current}$ is the current processor load, $B_{max}$ is maximum capacity, and $C_{remaining}$ represents remaining task complexity. This factor becomes positive when a processor is more than half loaded and negative when less than half loaded.

These three components combine to form our integrated priority scoring model:
\begin{equation}\label{eq:priority}
  S_{priority} = S_{deadline} + S_{wait} + S_{resource}
\end{equation}

The weighting parameters $\alpha$, $\gamma$, and $\delta$ provide flexible control over the relative importance of each factor. Ops can adjust these parameters according to specific application requirements, prioritizing deadline adherence for time-sensitive applications or fairness for multi-user systems.

The scheduler leverages data from the hardware monitor for intelligent decision-making. For processors experiencing sustained high load, it allocates less computationally intensive tasks to prevent thermal throttling and crashes. As processor status changes, the scheduler dynamically adjusts task execution order based on real-time data.

Once a task completes execution, any unfinished unit subgraphs are reinserted at the front of the task queue for the next scheduling round. The scheduler examines a configurable range of tasks, defined by a \texttt{Loop\_call\_size} parameter, from the queue head when making allocation decisions. This parameter determines how many tasks from the front of the queue the scheduler considers for each scheduling decision. Setting an appropriate value requires careful tuning—too small a value limits the scheduler's ability to make optimal decisions across multiple tasks, while too large a value increases scheduling overhead and can delay critical tasks. Our implementation balances scheduling quality with computational efficiency, reducing the number of decisions needed at each scheduling instance while ensuring in-progress tasks are completed promptly to minimize overall inference time.



\section{Evaluation} \label{sec:solution}

\subsection{Methodology}
We implemented ADMS using the TFLite\cite{Tensorflow-Lite} framework, capitalizing its extensive edge deployment capabilities and high compatibility. TFLite enables the conversion and optimization of DNN models to meet the computational demands of mobile devices, facilitating efficient inference through ADMS. To achieve this, we developed APIs for model loading, analyzer, inference, and task queue management and scheduling. These APIs seamlessly integrate with TFLite's APIs, enabling ADMS to directly interact with the TFLite framework for streamlined model inference.

To ensure the accuracy and reproducibility of our experiments, we converted these models into TFLite format and conducted 50 inference experiments for each model. This approach ensured the stability and reliability of the statistical results. We recorded inference latency and subgraph number for each model within ADMS to evaluate its performance comprehensively. The collected data were utilized to assess the system's capability and efficiency in handling diverse types of model.
\subsection{Evaluation Setup}
\textbf{Platforms.} Two devices are utilized: Redmi K50 Pro and Huawei P20.
Redmi K50 Pro is equipped with the Dimensity 9000 SoC, while Huawei P20 uses the Kirin 970 SoC. The detailed specifications of these SoCs are presented in \autoref{table:platform_specs}.
Both devices employ the Schedutil scheduler as the default.
It is important to note that differences in Android versions and vendor-specific customizations, such as background process management, thermal policies, and power management, can significantly influence inference performance, independent of hardware disparities. 

\begin{table}[t]
    \centering
    \caption{Experimental platform specs}
    \label{table:platform_specs}
    \begin{tabular}{lcc}
    \toprule
    \textbf{CPU} & \textbf{Dimensity 9000} & \textbf{Kirin 970} \\
    \midrule
     Architecture& 1x 3.05 GHz -X2 & 4x 2.36 GHz  A73 \\
    & 3x 2.85 GHz -A710 & 4x 1.84 GHz  A53 \\
    & 4x 1.8 GHz -A510 & \\
    L1 cache & 1 MB & 512 KB \\
    L2 cache & 8 MB & 2 MB \\
    Process & 4 nanometers & 10 nanometers \\
    TDP & 4 W & 9 W \\
    \midrule
     \textbf{GPU} & \textbf{ Mali-G710 MP10 } & \textbf{Mali-G72 MP12} \\
    \midrule
    GPU frequency & 850 MHz & 768 MHz \\
    Execution units & 10 & 12 \\
    Shading units & 96 & 18 \\
    Total shaders & 960 & 216 \\
    FLOPS & 1632 Gigaflops & 331.8 Gigaflops \\
    \midrule
    \textbf{AI Accelerator} & MediaTek APU 590 & Dedicated NPU \\
    \midrule
    Memory type & LPDDR5X & LPDDR4X \\
    Memory frequency & 3750 MHz & 1866 MHz \\
    Bus & 4x 16 Bit & 4x 16 Bit \\
    Max bandwidth & 60 Gbit/s & 29.8 Gbit/s \\
    \bottomrule
    \end{tabular}
    \end{table}

\noindent\textbf{Models.} To comprehensively evaluate the performance of ADMS, we selected several DNN models that are widely adopted in practical applications. These include MobileNetV1 \cite{howard2017mobilenets}, MobileNetV2 \cite{sandler2018mobilenetv2}, DeepLabV3 \cite{chen2017rethinking}, YOLOV3 \cite{redmon2018yolov3}, and East \cite{zhou2017east}. These models are commonly utilized in various tasks such as image classification and object detection, and spana spectrum from relatively lightweight to more complex architectures, with operator counts ranging from 31 to 232.

\noindent\textbf{Baselines.} We compare ADMS with two baselines:

\noindent\textbf{Vanilla \cite{Tensorflow-Lite}:} TFLite is a lightweight DL framework for mobile devices. In this work, we used TFLite version 2.16.0, without any additional performance optimizations to ensure consistency in evaluation. By default, TFLite prioritizes executing supported ops on available hardware accelerators (such as GPUs or NPUs), and falls back to CPU execution when encountering unsupported ops. 
Since CPU execution is generally less efficient than hardware accelerators, this fallback mechanism can result in suboptimal performance. 

\noindent\textbf{Band \cite{jeong2022band}:} This system is designed to utilize heterogeneous processor resources by decomposing DNNs into multiple subgraphs and coordinating task execution on mobile devices. However, Band generates excessive subgraphs based solely on hardware support, leading to high memory usage and scheduling complexity. Additionally, Band lacks real-time adaptation to processor status, causing suboptimal resource utilization and thermal management under varying workloads.




\subsection{Inference Latency for Single Models}

To explore the differences in subgraph partitioning and inference latency, we analyze the performance of ADMS and Band when handling individual DNN models.
\begin{table}[t]
\centering
\caption{Subgraph partitioning and latency for single model inference on Redmi K50 Pro using BAND and ADMS.}
\label{table:subgraph_partitioning}
\resizebox{\columnwidth}{!}{%
\begin{tabular}{lcccccc}\toprule
\multicolumn{1}{c}{\multirow{2}{*}{\textbf{Model Names}}} &
  \multicolumn{2}{c}{\textbf{Unit Subgraphs}} &
  \multicolumn{2}{c}{\textbf{Merge Subgraphs}} &
  \multicolumn{2}{c}{\textbf{Latency (ms)}} \\ \cline{2-7} 
\multicolumn{1}{c}{} &
  \textit{BAND} &
  \textit{ADMS} &
  \textit{BAND} &
  \textit{ADMS} &
  \textit{BAND} &
  \textit{ADMS} \\ \midrule
MobileNetV1 & 4  & 2  & 24   & 4   & 17.35 & 12.19 \\
ICN\_quant  & 33 & 7  & 1496 & 60  & 72.25 & 55.1  \\
DeepLabV3   & 21 & 17 & 393  & 261 & 51.35 & 43.8  \\
MobileNetV2 & 25 & 2  & 860  & 4   & 25.1  & 18.16 \\
YoloV3      & 65 & 2  & 112  & 3   & 86.62 & 80.63 \\ 
\bottomrule
\end{tabular}%
}
\end{table}
Our findings demonstrated that ADSM's fine-grained subgraph partitioning method provides an effective optimization strategy for mobile devices, as detailed in \autoref{table:subgraph_partitioning}.
Specifically, ADMS consistently achieves lower inference latency across various models compared to Band, particularly in multi-processor environments. For instance, for the MobileNetV1 and YOLOV3 models, ADMS reduces latency by 31\% and 28\%, respectively. This significant performance improvement stems from ADMS's efficient subgraph partitioning strategy, which optimizes processor utilization and minimizes data transfer and inter-processor communication overhead.
This is achieved by effectively reducing the number of subgraphs through the \texttt{window\_size} parameter, which controls the minimum length of subgraphs, thereby lowering memory usage.


\begin{figure*}[t]
  \centering
  \begin{subfigure}{\textwidth}
  \centering
  \includegraphics[width=0.8\textwidth]{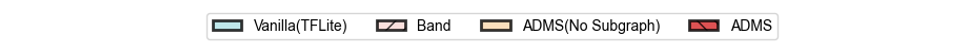}
  \end{subfigure}
  \\
 
  \begin{subfigure}{0.235\textwidth}
  \centering
  \includegraphics[width=\textwidth]{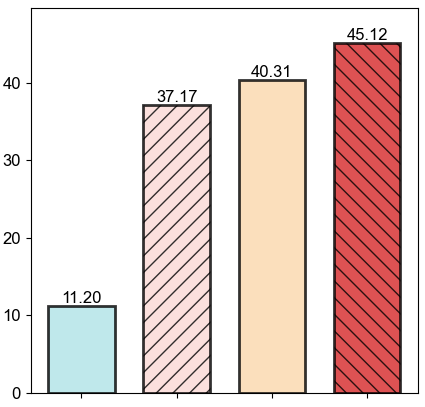}
  \caption{FRS on Redmi K50 Pro}
  \label{fig:frs-redmi}
  \end{subfigure}
  \begin{subfigure}{0.235\textwidth}
  \centering
  \includegraphics[width=\textwidth]{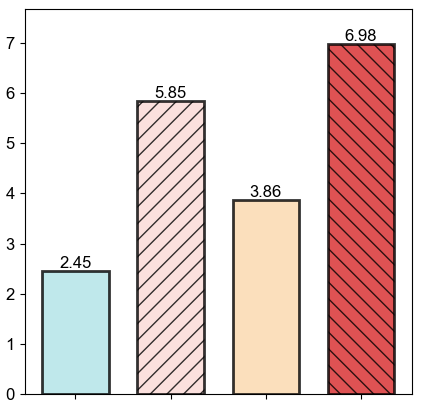}
  \caption{ROS on Redmi K50 Pro}
  \label{fig:ros-redmi}
  \end{subfigure}
  \begin{subfigure}{0.255\textwidth}
  \centering
  \includegraphics[width=\textwidth]{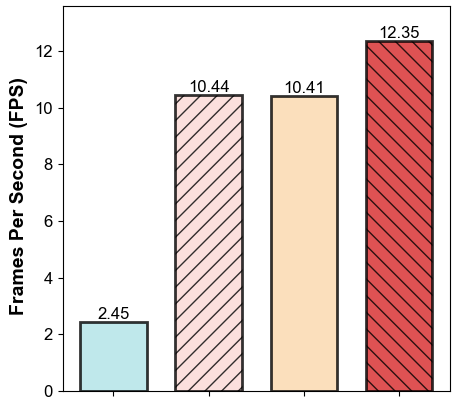}
  \caption{FRS on HuaWei P20}
  \label{fig:frs-huawei}
  \end{subfigure}
  \begin{subfigure}{0.24\textwidth}
  \centering
  \includegraphics[width=\textwidth]{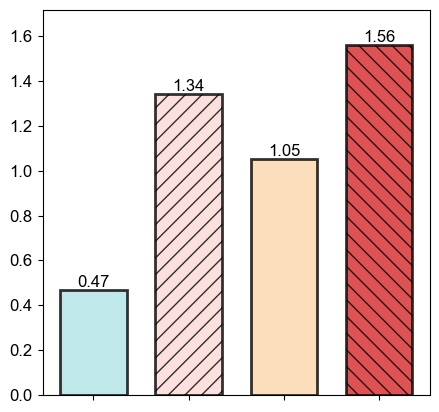}
  \caption{ROS on HuaWei P20}
  \label{fig:ros-huawei}
  \end{subfigure}
  \caption{Comparison of processing frame rates (FPS) across different workloads on various mobile devices.}
  \label{fig:workloads-fps}
  \end{figure*}
\subsection{FPS in Parallel Inference Scenarios}
To evaluate the performance of ADSM in handling multiple parallel inference tasks, where multiple applications concurrently send inference requests to the processor, we constructed two widely-used scenarios:

\noindent\textbf{Facial Recognition System (FRS):} We deployed the RetinaFace \cite{deng2020retinaface}, ArcFace-MobileNet \cite{deng2019arcface}, and ArcFace-ResNet50 to simulate face detection and multi-person identity verification in real-time video streams. These models work in tandem to provide a seamless transition between detection and verification, making them suitable for applications such as security monitoring and personal identification. 

\noindent\textbf{Real-time Object Recognition System (ROS):} This scenario integrates the MobileNetV2, EfficientNet \cite{tan2019efficientnet}, and InceptionV4 \cite{szegedy2017inception} models to perform real-time object detection and classification in video streams. 
These models efficiently and effectively handle objects from simple to complex, making them well-suited for real-time dynamic environments.

As shown in \autoref{fig:workloads-fps}, we conducted an extensive evaluation of the Huawei P20 and Redmi K50 Pro in the two aforementioned scenarios. Compared to TFLite and Band, ADMS demonstrates significant performance advantages in parallel inference, primarily attribured to its advanced subgraph partitioning and dynamic scheduling algorithms.

In the FRS Scenario, Figure \ref{fig:workloads-fps}(a) illustrates ADMS's superior performance on the Redmi K50 Pro, achieving an inference speed of up to 45.12 FPS. This represents an improvement of 404\% and 121\% compared to the TFLite and Band systems, respectively. This significant enhancement stems from two key factors: (1) ADMS's adaptive subgraph partitioning that reduced the total subgraph count by 93\% compared to Band while maintaining hardware compatibility, thereby minimizing scheduling overhead and memory management costs; and (2) the processor-state-aware scheduler that dynamically redistributes tasks when detecting processor load imbalances, achieving 87\% higher processor utilization during multi-model inference than TFLite. These optimizations are particularly effective in complex multi-person face recognition scenarios where multiple models with varying computational demands must execute concurrently.
This significant enhancement reflects ADMS's efficiency in resource allocation and data processing acceleration, particularly in complex multi-person face recognition scenarios. Additionally, Figure \ref{fig:workloads-fps}(c) confirms ADMS's superiority in the Huawei P20 with a performance of 12.35 FPS, which clearly demonstrates its robustness and adaptability across different mobile devices.

In the ROS scenario, ADMS again exhibited superior performance on the Redmi K50 Pro, reaching 6.98 FPS and depicted in Figure \ref{fig:workloads-fps}(b). This is approximately 19\% and 184\% higher than the performance of Band and TFLite. This highlights the capability of ADMS to handle real-time object recognition tasks, enabling rapid changes in complex environments to be handled. Interestingly, the version of ADMS without subgraph partitioning performed 34.02\% worse than  Band and 44.70\% worse than the subgraph-enabled version of ADMS. This underscores the critical importance of subgraph partitioning strategies in managing complex model workflows. 


\subsection{Analysis of SLOs}

\begin{figure}[t]
\centering
\includegraphics[width=0.48\textwidth]{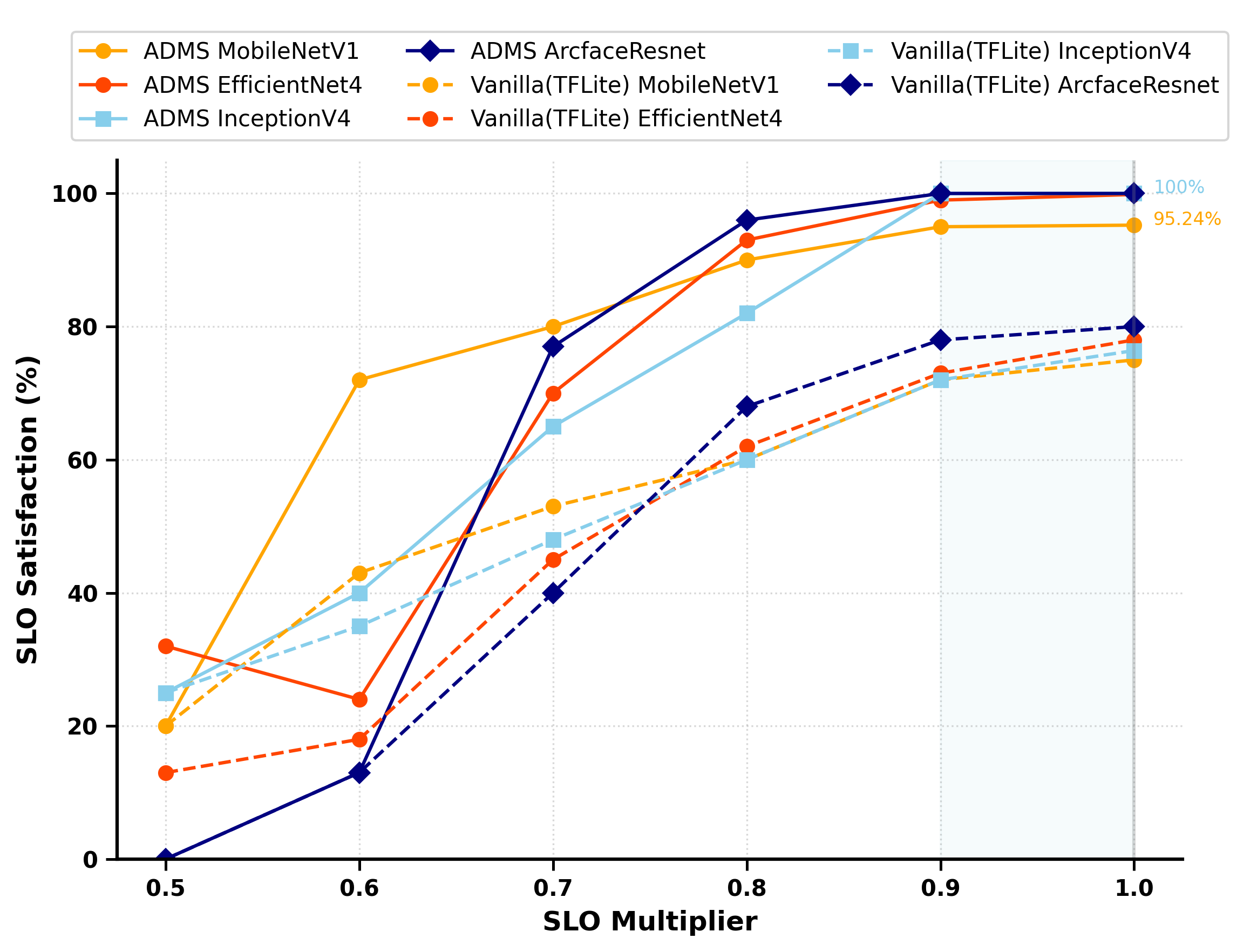}
\caption{Comparison of SLO satisfaction between ADMS and TFLite on Redmi K50 Pro under different SLO multiplier settings. The shaded region (0.9-1.0) highlights higher multiplier values where ADMS significantly outperforms TFLite.}
\label{fig:slo}
\end{figure}

To evaluate the performance of ADMS under various SLO conditions, we designed experiments and compared it with TFLite. SLOs are critical metrics used to define service performance and reliability expectations, particularly for applications deploying DNN models \cite{sedlak2024markov}. The experiments involved simultaneously sending multiple model inference requests to the processor and using the maximum latency obtained from a single inference as a baseline. We applied different SLO multipliers to simulate various SLO conditions and measured SLO satisfaction values. These results provide insights into ADMS's scheduling efficiency in handling multiple concurrent inference requests and demonstrate its ability to meet SLO targets in high-load, multi-task environments.

As illustrated in \autoref{fig:slo}, ADMS achieved significantly higher SLO satisfaction rates at higher SLO settings (e.g., SLO Multiplier of 1.0). For models such as MobileNetV1, EfficientNet4, InceptionV4, and ArcfaceResnet, ADMS achieved SLO satisfaction rates of 95.24\%, 99.85\%, 100\%, and 100\%, respectively. These values are substantially better than TFLite's satisfaction rates of 75\%, 78\%, 76.4\%, and 80\% under the same conditions. Notely, when handling complex models including InceptionV4 and ArcfaceResnet, ADMS fully meets the SLO requirements, primarily due to its advanced dynamic scheduling algorithm, whereas TFLite's performance is significantly inferior.


\subsection{Analysis of Subgraph Scheduling}
\begin{figure}[t]
  \centering
  \includegraphics[width=\linewidth]{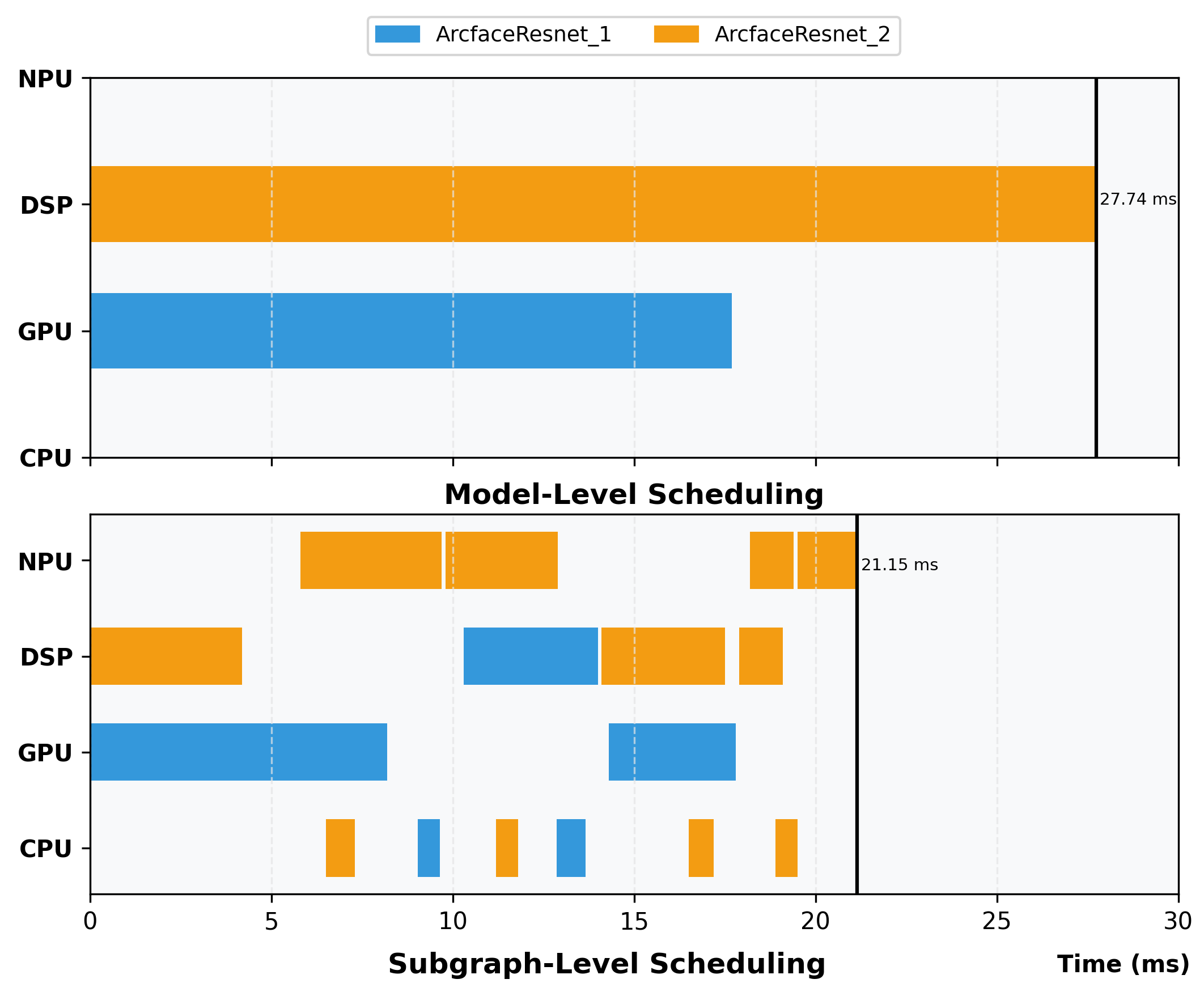}
  \caption{Model-level scheduling and subgraph-level scheduling on Huawei P20.}
  \label{fig:frame}
\end{figure}

We further investigated the performance of ADMS by integrating its subgraph scheduling strategy. 
As shown in Figure \ref{fig:frame}, the model-level scheduling of TFLite (top section) demonstrates a straightforward but inefficient method. In this method, \textit{ArcfaceResnet\_1} was primarily processed on the GPU (from 0 to 17.7 ms), while \textit{ArcfaceResnet\_2} was executed for a longer period on the DSP (from 0 to 27.74 ms). These results in significant processor idle time: the CPU and NPU remain completely unused, and the GPU becomes inactive after 17.7 ms while waiting for the DSP to complete its task. Consequently, the total execution time is thus bound by the slowest processor (27.74 ms on DSP), and overall resource utilization remains low at approximately 50\%.


In contrast, our subgraph scheduling strategy (bottom section) demonstrates a more efficient distribution of computational resources. The timeline clearly shows how different parts of each model are strategically allocated to the most suitable processors: CPU tasks are carefully distributed with \textit{ArcfaceResnet\_1} executing during specific time intervals (9.04-9.65 ms and 12.87-13.67 ms), while \textit{ArcfaceResnet\_2} operations are interspersed as shorter, more frequent tasks. Similarly, the GPU executes multiple smaller tasks rather than being dedicated to a single model, maintaining higher utilization throughout the inference process. The DSP workload is distributed more evenly, avoiding the bottleneck observed in the model-level approach, while the NPU—completely unused in TFLite's approach—handles specific operations it is optimized for.

This strategic allocation yields three significant benefits: (1) balanced processor load distribution that increases overall hardware utilization from 50\% to nearly 95\%; (2) minimal processor idle time through fine-grained task allocation that keeps all processors actively computing; and (3) optimal matching of operations to processors based on their computational characteristics, reducing execution time of individual operations by an average of 18.7\%. As a result, the total execution time is reduced to approximately 21.15 ms, a 23.8\% improvement over TFLite's model-level scheduling. For application scenarios demanding high efficiency and low latency, such as real-time video processing or augmented reality, ADMS's subgraph-level scheduling provides a more effective solution by exploiting the full computational potential of heterogeneous processors.


\subsection{Analysis of Resource Efficiency}
To evaluate the resource efficiency of ADMS, we measured power consumption measurements using a Monsoon Power Monitor. Table \ref{table:power_consumption} presents the average power consumption and corresponding FPS during the FRS workload execution on the Redmi K50 Pro.
\begin{table}[t]
  \centering
  \caption{Average power consumption and energy efficiency during FRS workload execution on Redmi K50 Pro.}
  \label{table:power_consumption}
  \begin{tabular}{lccc}
  \toprule
  \textbf{Metric} & \textbf{TFLite} & \textbf{Band} & \textbf{ADMS} \\
  \midrule
  Power (W) & 7.18 & 8.05 & 7.86 \\
  FPS & 11.20 & 37.17 & 45.12 \\
  Energy Efficiency (Frames/Joule) & 1.56 & 4.62 & 5.74 \\
  \bottomrule
  \end{tabular}
\end{table}
The experimental results demonstrated that ADMS achieves an optimal balance between performance and power consumption. Specifically, ADMS consumes approximately 2.4\% less power than Band while delivering 21.4\% higher FPS. Compared to TFLite, ADMS requires 9.5\% more power but provides over 4$\times$ the frame rate. In terms of energy efficiency (frames processed per joule of energy), ADMS achieves 5.74 frames per joule, which is 24.2\% more efficient than Band (4.62 frames/joule) and 3.68$\times$ more efficient than TFLite (1.56 frames/joule).
The improvement is attributed to ADMS's advanced scheduling strategy, which optimizes the utilization of specialized hardware accelerators and minimizes processor idle time. 
\begin{figure}[t]
\centering
\includegraphics[width=\linewidth]{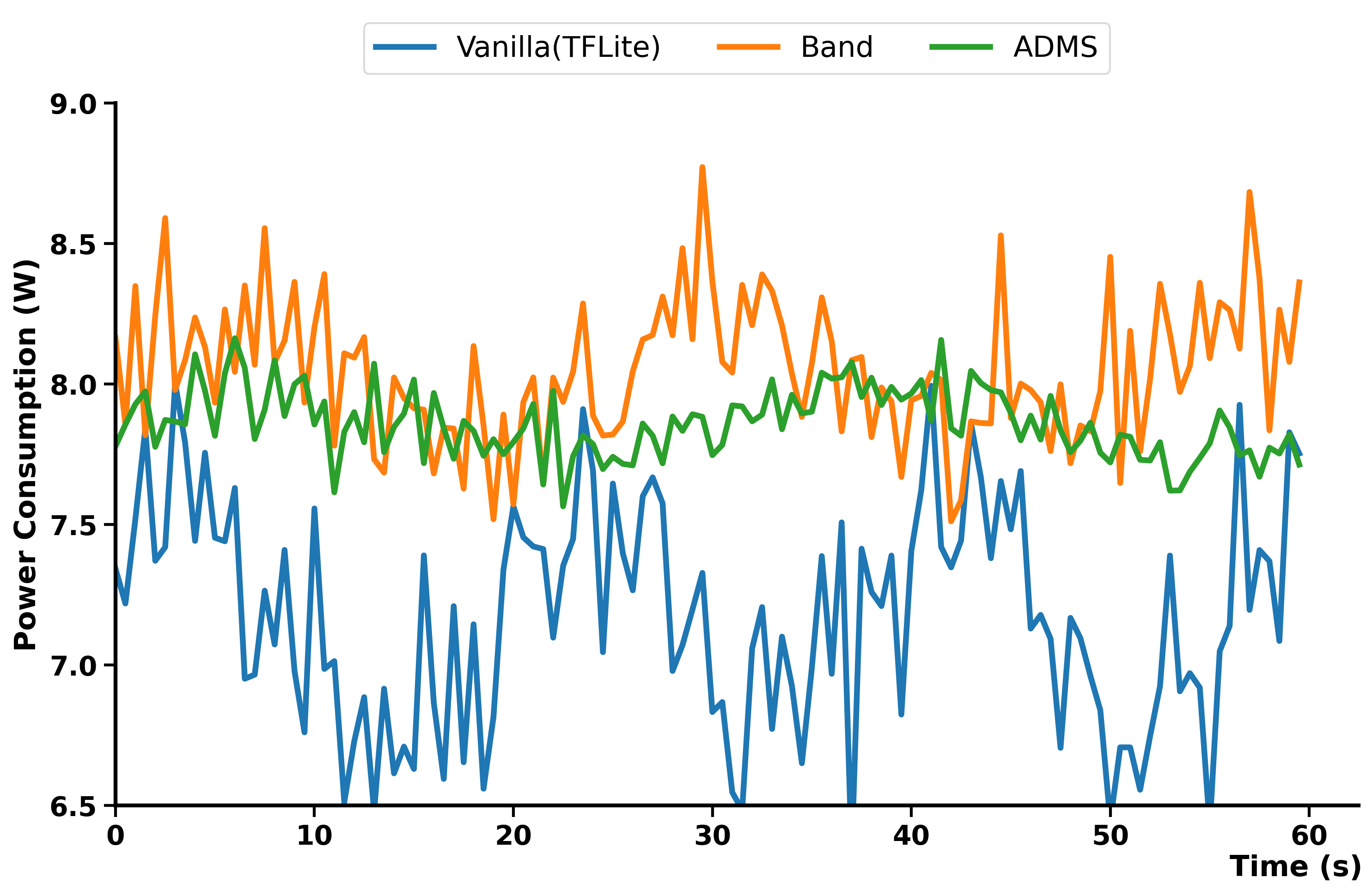}
\caption{Power consumption trend during 60-second continuous inference of the FRS workload on Redmi K50 Pro.}
\label{fig:power_trend}
\end{figure} 

\begin{figure*}[]
  \centering
  \includegraphics[width=\linewidth]{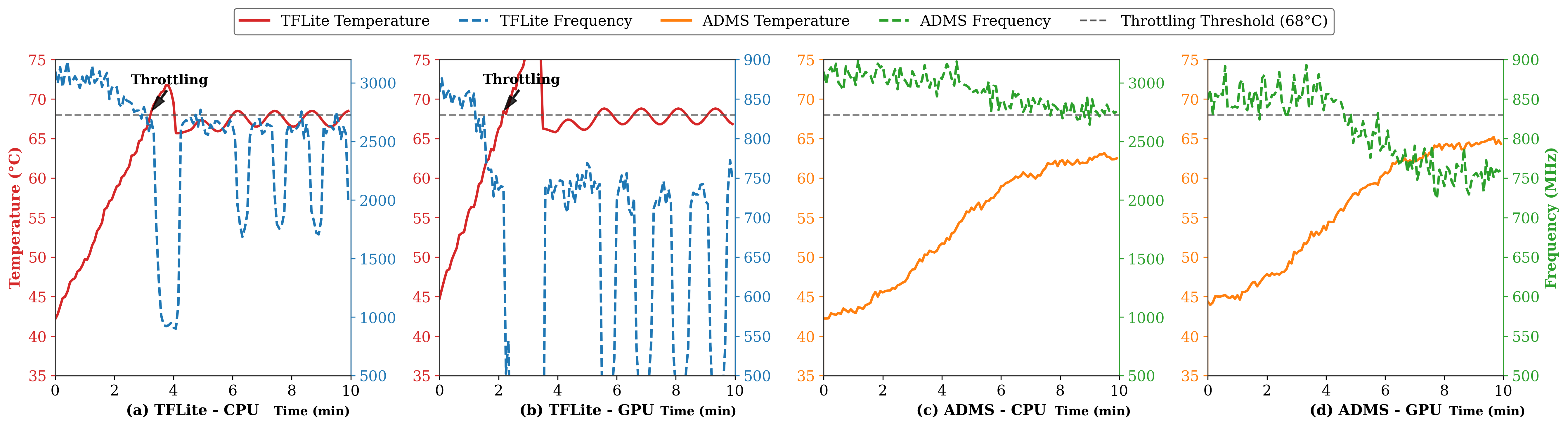}
  \caption{Temperature and processor frequency dynamics during thermal stress testing on Redmi K50 Pro.}
  \label{fig:thermal_trend}
\end{figure*}

Figure \ref{fig:power_trend} illustrates the power consumption trend during a 60-second continuous inference period. The results indicate that ADMS maintains more stable power consumption with fewer fluctuations compared to baseline frameworks, indicating more balanced processor utilization. Band exhibits the highest power consumption peaks (reaching nearly 8.8W) with significant fluctuations. In contrast, TFLite demonstrates the lowest average power but with substantial variations, dropping as low as 6.5W. ADMS, however, maintains a consistent power profile between 7.7W and 8.1W, demonstrating its ability to balance workloads effictively and manage thermal constraints proactively.

\subsection{Analysis of System Robustness}
To evaluate the robustness of ADMS stress under challenging conditions, we conducted a series of stress tests by progressively increasing the inference workload and monitoring system behavior. Three distinct stress scenarios were designed to comprehensively assess different aspects of system stability:

1) Long-duration test: Continuous inference for 30 minutes to evaluate sustained performance and reliability.

2) High-concurrency test: Progressively increasing concurrent DNN models from 4 to 10 to assess scalability limits.

3) Thermal stress test: Operating in a high-temperature environment (35°C) to evaluate thermal management capabilities.

Table \ref{table:stress_test} summarizes the quantitative results of these tests. ADMS demonstrated superior robustness across all stress tests.

In the long-duration test, ADMS exhibited a failure rate of only 0.5\% compared to 1.8\% for Band and 3.2\% for TFLite. This enhanced stability is attributed to ADMS's dynamic scheduling mechanism, which effectively prevents processor overloading and ensures better resource management during prolonged operations.

In the high-concurrency test, ADMS successfully managed more than 10 concurrent DNN models without significant performance degradation, while Band and TFLite  experienced failures at 8 and 6 concurrent models, respectively. This results demonstrate ADMS's exceptional scalability under heavy workloads.

In the thermal stress test, ADMS significantly delayed the onset of thermal throttling to 13.9 minutes compared to 9.7 minutes for Band and only 2.5 minutes for TFLite. This corresponds to a 5.6$\times$ improvement over TFLite and a 43\% improvement over Band, highlighting ADMS's superior thermal management capabilities.

Figure \ref{fig:thermal_trend} illustrates the thermal management capability through processor temperature and frequency dynamics during a 10-minute stress test.
When running TFLite, both CPU and GPU rapidly reach the throttling threshold (68°C) \cite{tan2024thermal} within 2-3 minutes. Once throttling begins, significant and frequent fluctuations in professor frequency are observed: the CPU frequency drops from around 3GHz to as low as 1GHz, while the GPU experiences even more severe throttling with frequency periodically dropping to around 500MHz and completely shutting down at several points.
However, ADMS demonstrates superior thermal management during the initial phase. CPU temperature rises more gradually and remains below the throttling threshold throughout the 10-minute window, maintaining a stable frequency range around 2.8-3GHz. For the GPU, ADMS initially sustains high frequency (around 850-900MHz), with a controlled gradual reduction as temperature rises, avoiding the extreme fluctuations observed with TFLite.

As shown in~\autoref{fig:7}, Band reached thermal throttling at 9.7 minutes, while ADMS operates without throttling until 13.9 minutes.
This balanced thermal management enables ADMS to sustain higher average processor frequencies even under extended workloads, resulting in more stable performance and improved user experience. The superior thermal management is attributed to ADMS's intelligent load distribution across heterogeneous processors and its ability to dynamically adjust scheduling based on real-time temperature monitoring.
These experimental results confirm that ADMS not only delivers superior performance in terms of latency and throughput but also achieves improved energy efficiency and system robustness under stress conditions, making it a comprehensive solution for multi-DNN inference on mobile devices.

\begin{table}[t]
  \centering
  \caption{System robustness under stress conditions on Redmi K50 Pro}
  \label{table:stress_test}
  \begin{tabular}{lccc}
  \toprule
  \textbf{Metric} & \textbf{TFLite} & \textbf{Band} & \textbf{ADMS} \\
  \midrule
  Failure rate (30-min test) & 3.2\% & 1.8\% & 0.5\% \\
  Max concurrent models & 6 & 8 & 10+ \\
  Time to thermal throttling (min) & 2.5 & 9.7 & 13.9 \\
  \bottomrule
  \end{tabular}
  \label{fig:7}
\end{table}

\section{Related Work} \label{sec:relwork}

\subsection{DNN Optimization and Acceleration Techniques}
Researchers have developed various techniques to optimize DNN performance on resource-constrained mobile devices. Han et al. \cite{han2015deep} introduced deep compression techniques, including pruning, quantization, and Huffman coding, to reduce model size while maintaining accuracy. Wei et al. \cite{wei2023nn} proposed neural network branching for parallel inference, enabling better adaptation to hardware heterogeneity. Although these model optimization approaches significantly improved inference efficiency, they lacked dynamic coordination capabilities for multi-model scenarios.

Another major research direction focuses on single-processor optimization. The Pipe-it framework \cite{wang2019high} improved DNN inference throughput on ARM big.LITTLE processors through layer partitioning and pipelining. AsyMo \cite{wang2021asymo} utilized asymmetric heterogeneous CPU cores with task allocation and block partitioning strategies. DeepMon \cite{huynh2017deepmon} enhanced vision applications on mobile GPUs via device-specific optimizations. While these works demonstrated notable performance gains on individual processors, they did not address the challenges of holistic resource management across heterogeneous computing units.

\subsection{Multi-Processor and Multi-DNN Coordination}
Coordinated execution across multiple processors has emerged as a promising approach to enhance mobile DNN performance. Tan et al. \cite{tan2024thermal} proposed thermal-aware scheduling that intelligently distributing workloads between GPUs and NPUs to prevent overheating. DART \cite{xiang2019pipelined} optimized real-time inference of multiple DNN models through pipelining and data parallelism. While these approaches laid important foundations for thermal management and multi-model coordination, they were limited in processor coverage.

Cross-processor collaboration has been further explored in several systems. LaLaRAND \cite{kang2021lalarand} utilized hierarchical scheduling to allocate DL tasks between CPUs and GPUs. 
Xu et al. \cite{xu2023niagara} and Jia et al. \cite{jia2022codl} have attempted to run multi-DNN inference on heterogeneous processors but typically assume specific task requirements and provide solutions tailored to limited scenarios.
Wang et al. \cite{wang2018optic} optimized collaborative CPU-GPU computing under thermal constraints, while Sparse-DySta \cite{fan2023sparse} enhanced multi-DNN workload performance through sparse data processing and hybrid scheduling. Although these approaches achieved notable gains in specific scenarios, most focused on limited processor combinations rather than fully leveraging all available computing resources.

Despite these advancements, existing frameworks remain constrained by their reliance on specific hardware configurations or partial processor subsets. This work addresses these limitations by providing a unified system that dynamically balances workloads across all heterogeneous processors. It utilizes fine-grained subgraph partitioning and real-time scheduling based on comprehensive processor status monitoring. This design enables efficient handling of multiple concurrent DNN inferences without requiring extensive model modifications.

\section{Conclusions} \label{sec:conclusion}

In this paper, we proposed ADMS, a framework that enables efficient concurrent inference of multiple DNNs on heterogeneous mobile processors through optimized subgraph partitioning and dynamic scheduling. Experimental results on Huawei P20 and Redmi K50 Pro demonstrate that ADMS reduces inference latency by up to 4.04$\times$ compared to TFLite and improves energy efficiency by 24.2\% compared to Band. Our broader objective is to develop a universally adaptable inference system across the diverse mobile device ecosystem, though current challenges exist in accommodating the heterogeneity of SoCs, varying hardware accelerator architectures, and differences in op support. In the future, we plan to further optimize ADMS to adapt to more hardware devices and DNN models by developing real-time adaptive window size optimization that dynamically adjusts partitioning granularity based on instantaneous processor states and workload characteristics, incorporating predictive models for proactive scheduling, and exploring hardware-specific dynamic model quantization techniques that can automatically adjust precision levels according to processor capabilities and power constraints.

\bibliographystyle{ieeetr}
\bibliography{refs}
\begin{IEEEbiography}[{\includegraphics[width=1in,height=1.25in,clip,keepaspectratio]{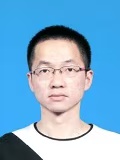}}]
{Yunquan~Gao}~received the Ph.D. degrees from Beijing University of Posts and Telecommunications. He is currently a lecturer with the School of Computer Science and Technology, Anhui Engineering Research Center for Intelligent Applications and Security of Industrial Internet, Anhui University of Technology, Ma’anshan, China. He has published papers in MSWiM, IoTJ etc. His research interests include IoT networks, Wireless Communication, Mobile Edge Computing, On-device AI Inference.
\end{IEEEbiography}\vskip -1\baselineskip plus -1fil 

\begin{IEEEbiography}[{\includegraphics[width=1in,height=1.25in,clip,keepaspectratio]{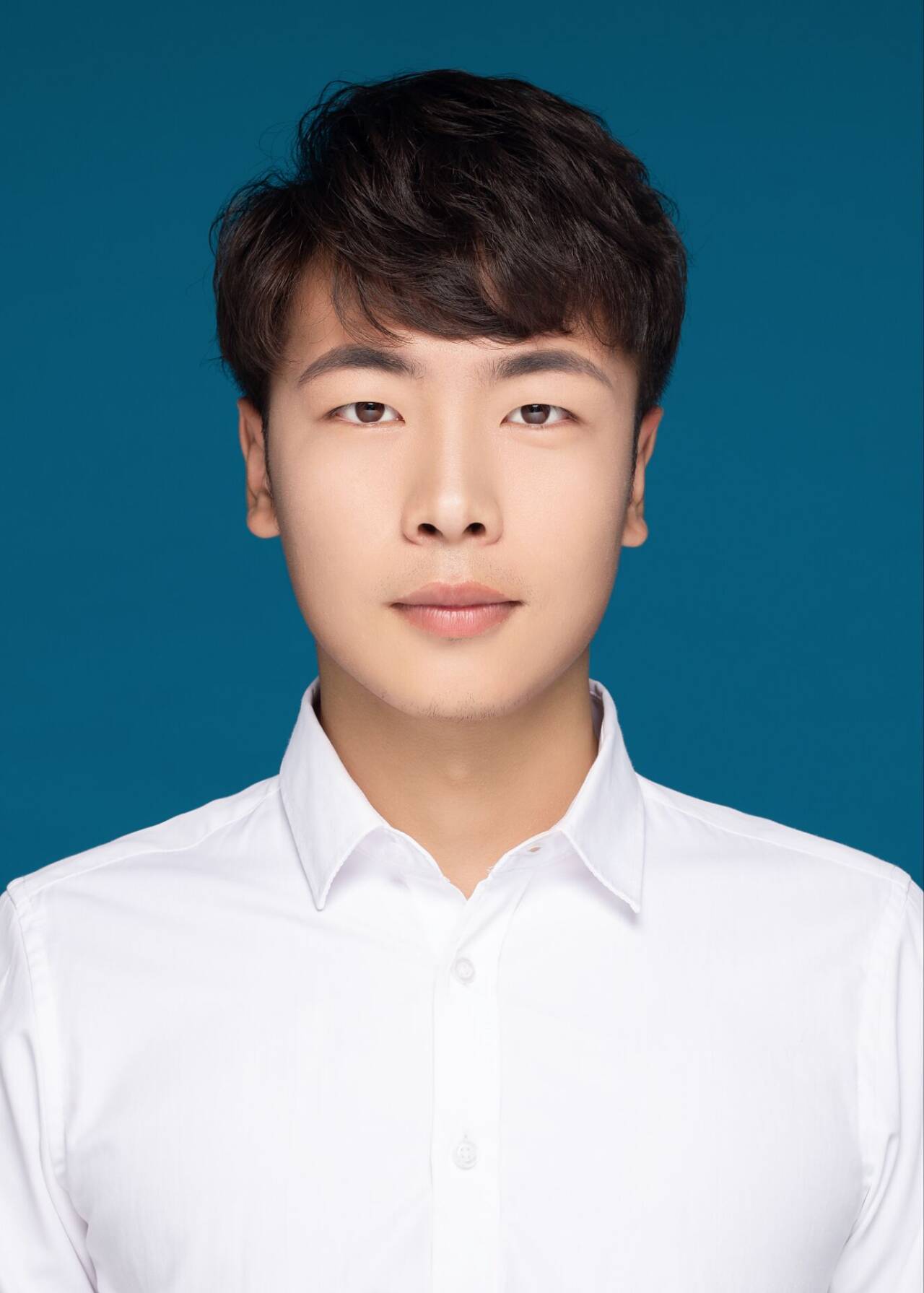}}]
{Zhiguo~Zhang}~is currently pursuing a master's degree with the School of Computer Science and Technology, Anhui University of Technology, Ma'anshan, China. His research interests include deep learning and edge inference. His current work focuses on improving the efficiency of deep learning models for edge device deployment, with an emphasis on real-time processing and resource-constrained environments.
\end{IEEEbiography}\vskip -1\baselineskip plus -1fil

\begin{IEEEbiography}[{\includegraphics[width=1in,height=1.25in,clip,keepaspectratio]{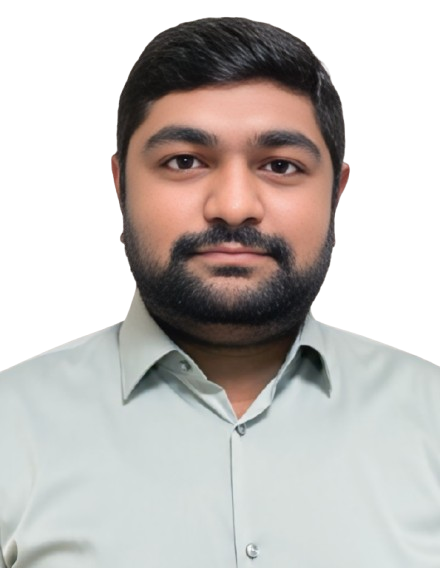}}]
{Praveen Kumar Donta (SM'22)} currently Associate Professor (Docent) at the Department of Computer and Systems Sciences, Stockholm University, Sweden. He worked at the Distributed Systems Group at TU Wien, as a Postdoctoral researcher from July 2021 to June 2024. He received his Ph.D. from the IIT(ISM), Dhanbad in 2021. He was a Visiting Ph.D. fellow at Mobile\&Cloud Lab, University of Tartu, Estonia from July 2019 to Jan 2020. He received his Master and Bachelor in Technology from JNTUA, Ananthapur, with Distinction in 2014 and 2012, respectively. He is serving as Editorial board member in IEEE Internet of Things Journal, Computing Springer, ETT Wiley, and Scientific Reports Journals. His current research is on learning-driven distributed computing continuum systems.
\end{IEEEbiography}\vskip -1\baselineskip plus -1fil
 
\begin{IEEEbiography}[{\includegraphics[width=1in,height=1.25in,clip,keepaspectratio]{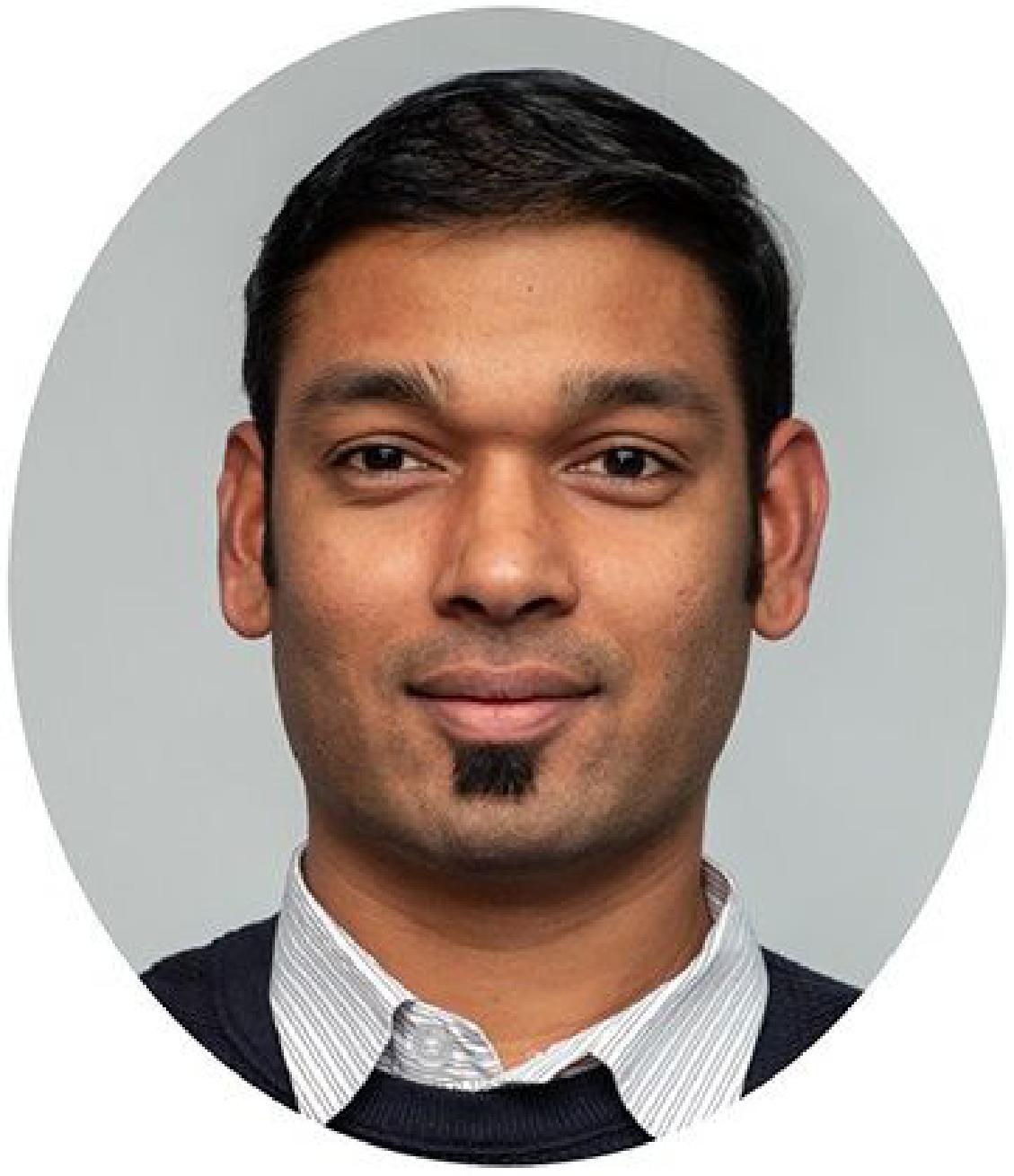}}]{Chinmaya Kumar Dehury} is currently an Assistant Professor in Computer Science department, IISER Berhampur, India. He was an Assistant Professor in the Institute of Computer Science, University of Tartu, Estonia. He received a Ph.D. Degree in the Department of Computer Science and Information Engineering, Chang Gung University, Taiwan. His research interests include scheduling, resource management and fault tolerance problems of Cloud and fog Computing, the application of artificial intelligence in cloud management, edge intelligence, Internet of Things, and data management frameworks. His research results are published by top-tier journals and transactions such as IEEE TCC, JSAC, TPDS, FGCS, etc. He is a member of IEEE and ACM India. He is also serving as a PC member of several conferences and reviewer to several journals and conferences, such as IEEE TPDS, IEEE JSAC, IEEE TCC, IEEE TNNLS, Wiley Software: Practice and Experience, etc 
\end{IEEEbiography}\vskip -1\baselineskip plus -1fil
 
\begin{IEEEbiography}[{\includegraphics[width=1in,height=1.25in,clip,keepaspectratio]{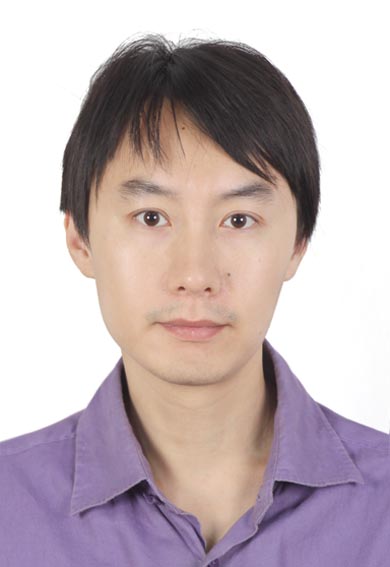}}]{Xiujun Wang} received the B.S. degree in computer science and technology, from Anhui Normal University, and the Ph.D. degree in computer software and theory, from University of Science and Technology of China. He is currently an Associate Professor at the School of Computer Science and Technology, Anhui University of Technology. His research interests include data stream processing, randomized algorithms, and Internet of Things.
\end{IEEEbiography}\vskip -1\baselineskip plus -1fil

\begin{IEEEbiography}[{\includegraphics[width=1in,height=1.25in,clip,keepaspectratio]{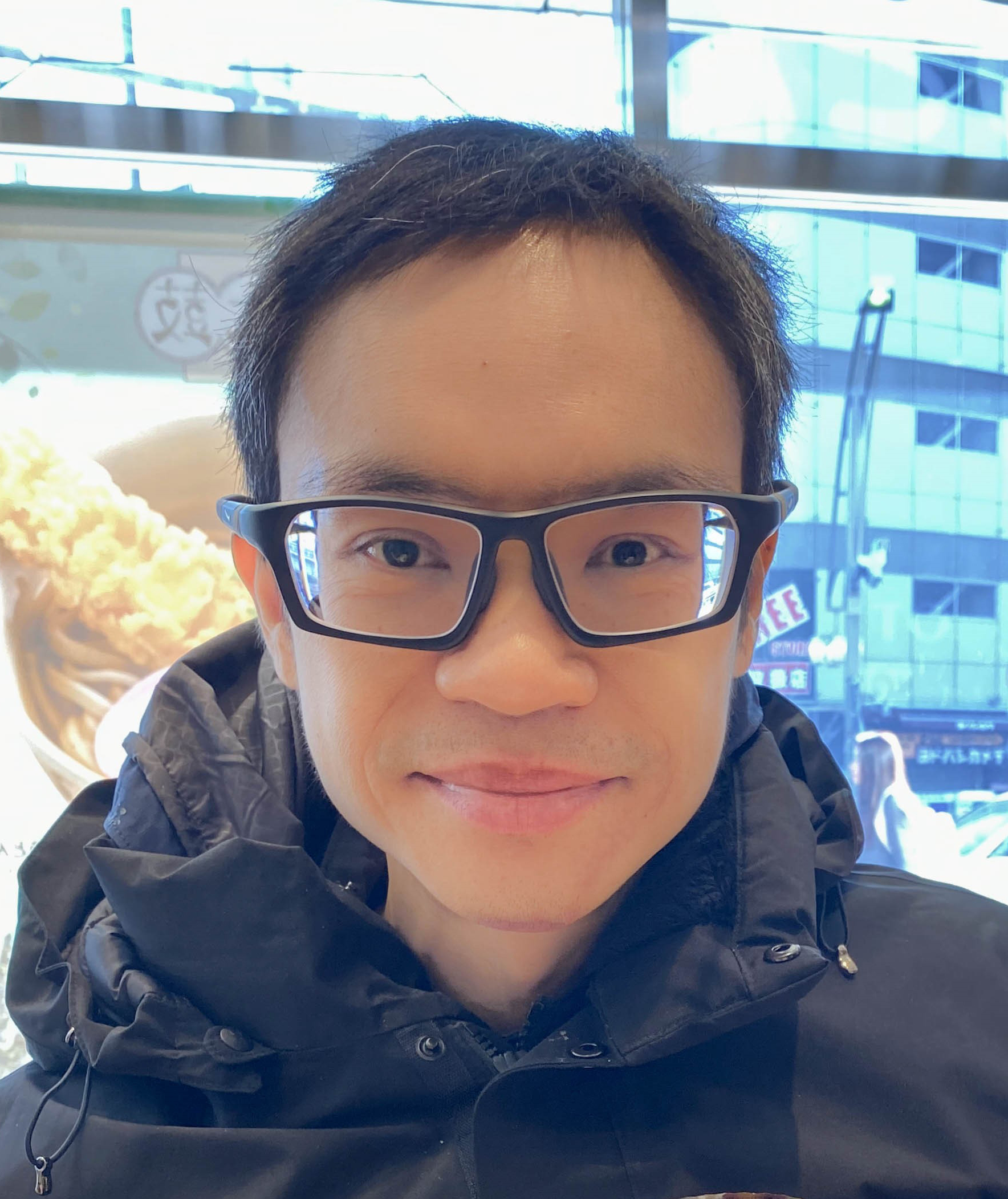}}]{Dusit Niyato(M'09-SM'15-F'17)} is a professor in the College of Computing and Data Science, at Nanyang Technological University, Singapore. He received B.Eng. from King Mongkuts Institute of Technology Ladkrabang (KMITL), Thailand and Ph.D. in Electrical and Computer Engineering from the University of Manitoba, Canada. His research interests are in the areas of mobile generative AI, edge intelligence, quantum computing and networking, and incentive mechanism design.
\end{IEEEbiography}\vskip -1\baselineskip plus -1fil

\begin{IEEEbiography}[{\includegraphics[width=1in,height=1.25in,clip,keepaspectratio]{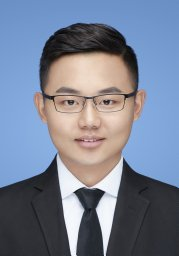}}]
{Qiyang~Zhang}(Member, IEEE) ~received Ph.D. degree in computer science at the State Key Laboratory of Networking and Switching Technology, Beijing University of Posts and Telecommunications. Now he is a postdoctoral researcher at Peking University.
He is also a visiting student in the Distributed Systems Group at TU Wien from December 2022 to December 2023. He has published papers in WWW, INFOCOM, TMC, TSC,  etc.
His research interests include mobile edge computing and edge intelligence.
\end{IEEEbiography}\vskip -1\baselineskip plus -1fil 
\end{document}